\documentclass[showpacs, superscriptaddress, showpacs, letterpaper, showkeys,
preprintnumbers, altaffilletter, amssymb, amsmath, amsfonts, prd,
twocolumn, floatfix, nofootinbib]{revtex4-1}

\usepackage{graphics}
\usepackage{color}
\usepackage{leftidx}
\usepackage[colorlinks=true]{hyperref}
\usepackage[caption=false]{subfig}

\def\gw#1{gravitational wave#1 (GW#1)\gdef\gw{GW}}

\newcommand{\diff}{{\mathrm d}}

\newcommand{\msun}{{\mathrm M}_{\odot}}

\def\mnras{\ref@jnl{MNRAS}}

\def\imbh#1{intermediate mass black hole#1(IMBH#1)\gdef\imbh{IMBH}}
\def\smbh#1{supermassive black hole#1(SMBH#1)\gdef\smbh{SMBH}}
\def\bbh#1{binary black hole#1 (BBH#1)\gdef\bbh{BBH}}
\def\bh#1{black hole#1 (BH#1)\gdef\bh{BH}}
\def\ns#1{neutron star#1 (NS#1)\gdef\ns{NS}}
\def\gw#1{gravitational wave#1 (GW#1)\gdef\gw{GW}}
\def\sn#1{core-collapse supernova#1 (CCSN#1)\gdef\sn{CCSN}}
\def\pnw#1{post-Newtonian#1 (PN#1)\gdef\pnw{PN}}
\def\eos#1{equation of state#1 (EoS#1)\gdef\eos{EoS}}
\def\grb#1{gamma-ray burst#1 (GRB#1)\gdef\grb{GRB}}
\def\amr#1{adaptive mesh refinement#1 (AMR#1)\gdef\amr{AMR}}
\def\isco#1{innermost stable circular orbit#1 (ISCO#1)\gdef\isco{ISCO}}
\def\cwb#1{Coherent WaveBurst#1 (CWB#1)\gdef\cwb{CWB}}

\usepackage{soul} 
\usepackage{ulem} 

\begin{document}

\title{Prospects For High Frequency Burst Searches Following Binary Neutron Star
Coalescence With Advanced Gravitational Wave Detectors}

\author{J. Clark}
\affiliation {University of Massachusetts Amherst, Amherst, MA 01003, USA } 
\author{A. Bauswein}
\affiliation{Aristotle University of Thessaloniki}
\author{L. Cadonati}
\affiliation {University of Massachusetts Amherst, Amherst, MA 01003, USA } 
\affiliation {Cardiff University, Cardiff, CF24 3AA, United Kingdom }
\author{H.-T. Janka}
\affiliation{Max Planck Institute For Astrophysics}
\author{C. Pankow}
\affiliation {University of Wisconsin--Milwaukee, Milwaukee, WI 53201, USA }
\author{N. Stergioulas}
\affiliation{Aristotle University of Thessaloniki}

\date{\today}

\begin{abstract}
The equation of state  plays a critical role in the physics of the merger of two
neutron stars.  Recent numerical simulations with microphysical equation of
state suggest the outcome of such events depends on the mass of the neutron
stars.  For less massive systems, simulations favor the formation of a
hypermassive, quasi-stable neutron star, whose oscillations produce a short,
high frequency burst of gravitational radiation.  Its dominant frequency content
is tightly correlated with the radius of the neutron star, and its measurement
can be used to constrain the supranuclear equation of state. In contrast, the
merger of higher mass systems results in prompt gravitational collapse to a
black hole.  We have developed an algorithm which combines waveform
reconstruction from a morphology-independent search for gravitational wave
transients with Bayesian model selection, to discriminate between  post-merger
scenarios and accurately measure the dominant oscillation frequency. We
demonstrate the efficacy of the method using a catalogue of simulated binary
merger signals in data from LIGO and Virgo, and we discuss the prospects for
this analysis in advanced ground-based gravitational wave detectors.  From the
waveforms considered in this work and assuming an optimally oriented source, we
find that the post-merger neutron star signal may be detectable by this
technique to $\sim 10\text{--}25$\,Mpc.  We also find that we successfully
discriminate between the post-merger scenarios with $\sim 95\%$ accuracy and
determine the dominant oscillation frequency of surviving post-merger neutron
stars to within $\sim 10$\,Hz, averaged over all detected signals.   This leads
to an uncertainty in the estimated radius of a non-rotating 1.6\,M$_{\odot}$
reference neutron star of $\sim 100\,$m.
\end{abstract}

%

\maketitle
\section{Introduction} 
\label{sec:intro} 
The inspiral and merger of binary neutron star systems (BNS) is one of the most
promising sources of \gw{s} for the second generation of ground-based detectors,
which include the US-based Advanced Laser Interferometer Gravitational Wave
Observatory (aLIGO)~\cite{2010CQGra..27h4006H}, the French-Italian Advanced
Virgo (AdV) observatory~\cite{2011CQGra..28k4002A,virgo_baseline} and the
Japanese Kamioka Gravitational Wave Detector (KAGRA)
observatory~\cite{2010CQGra..27h4004K}.  It is expected that the aLIGO-AdV
network will reach design sensitivity in 2018-2020~\cite{2013arXiv1304.0670L},
leading to the observation of $0.4\text{--}400$ BNS coalescence events per year
of operation~\cite{ratesPaper}, where the range in values are set by
uncertainties on the BNS coalescence rate.

The internal composition and properties of matter at supranuclear densities is
currently poorly understood and the equation of state (EoS) is not well
constrained~\cite{2007PhR...442..109L}.  The \gw{} signal from a BNS coalescence
carries important information on the EoS and offers and unprecedented
opportunity to probe the neutron star interior.  As the stars grow closer,
increasing tidal interactions imprint a distinctive EoS signature on the phase
evolution of the \gw{}
waveform~\cite{2008PhRvD..77b1502F,2010PhRvL.105z1101B,2012PhRvD..86d4030B}.
These tidal effects on the inspiral portion of the waveform may be detectable to
distances $\sim 100$\,Mpc in aLIGO, leading to the determination of NS radii to
an accuracy of about 1\,km~\cite{2009PhRvD..79l4033R}.   Complementary and
independent contraints on the EoS may be accessible from the post-merger part of
the coalescence signal.

The most likely post-merger scenario is the formation of a massive ($M>2\msun$),
differentially rotating neutron star, hereafter referred to as the post-merger
neutron star
(PMNS)~\cite{1994PhRvD..50.6247Z, 1996A&A...311..532R, 2005PhRvL..94t1101S,
2005PhRvD..71h4021S, 2007A&A...467..395O, 2007PhRvL..99l1102O,
2008PhRvD..77b4006A, 2008PhRvD..78b4012L, PhysRevD.78.084033,
2009PhRvD..80f4037K, 2011MNRAS.418..427S, 2011PhRvD..83d4014G,
2011PhRvD..83l4008H, 2011PhRvL.107e1102S, 2012PhRvL.108a1101B,
2012PhRvD..86f3001B, 2013PhRvL.111m1101B, 2013PhRvD..88d4026H,
2013arXiv1311.4443B, 2014arXiv1403.5672T, 2000ApJ...528L..29B}.
The stability of the PMNS against gravitational collapse depends on its mass.
Less massive systems result in a long-lived, stable PMNS.  For more massive
systems, or where insufficient material has been ejected during the merger,
centrifugal and thermal effects result in a quasi-stable remnant which
eventually undergoes gravitational collapse due to redistribution of energy and
angular momentum via viscous processes, radiation of \gw{s} and emission of
neutrinos (``delayed collapse'').  Sufficiently high-mass systems will result in
prompt collapse to a black hole (BH), emitting a high-frequency ring-down \gw{}
signal at $\sim$\,6--7\,kHz.  The detection of these stellar-mass black hole
ringdowns will be very challenging in the next generation of ground based \gw{}
detectors, due to their reduced sensitivity at high frequency; we will not
consider them further in this discussion.  We note that
in~\cite{2011PhRvD..83l4008H} the authors suggest two subclasses of the delayed
collapse scenario characterized by the lifetime of the post-merger remnant.  In
this work, however, we do not distinguish between the cases of long- and
short-lived PMNS.  Instead, we restrict our classification scheme to the two
cases: i) prompt collapse to a BH and ii) PMNS formation.  For simplicity, we
will hereafter refer to (ii) as `delayed collapse'. This term is also supposed
to subsume cases which actually do not lead to a gravitational collapse at all
because the PMNS is stable. Observationally, this scenario cannot be
distinguished from a true delayed collapse by the \gw{} signal immediately
following the merger.  Moreover, for some binary setups it may be very difficult
to decide based on numerical simulations whether the resulting PMNS is stable or
may eventually collapse because this would require long-term simulations which
also take into account the relevant physics of the secular evolution of the
PMNS.  Such simulations are currently unavailable.

In case that the PMNS survives prompt-collapse, transient non-axisymmetric
deformations in the post-merger remnant lead to a short duration ($\sim
10\text{--}100$\,ms) burst of \gw{s} which typically resembles an
amplitude-modulated, damped sinusoid with a dominant oscillation frequency $\sim
2\text{--}4$\,kHz associated with quadrupole oscillations in the fluid. In
addition to the dominant oscillations,  nonlinear couplings between certain
oscillation modes have been identified, which appear as secondary peaks in the
GW spectra~\cite{2011MNRAS.418..427S}. The spectral properties of this signal
carry a distinct signature of the EoS.  In ref.~\cite{2012PhRvD..86f3001B}, for
example, the authors perform a large scale survey using a wide variety of
different EoSs and establish the following correlation between the peak
frequency $f_{\mathrm{peak}}$ of the post-merger signal from 1.35-1.35$\msun$
binaries and the radius of a fiducial, non-rotating neutron star with mass
$1.6\,\msun$, $R_{1.6}$:
\begin{equation}\label{eq:fpeak_radius}
f_{\mathrm{peak}} = 
\begin{cases}
-0.2823 R_{1.6} + 6.284 &\mbox{for } f_{\mathrm{peak}} < 2.8\,{\mathrm{kHz}} \\
-0.4667 R_{1.6} + 8.713 &\mbox{for } f_{\mathrm{peak}} > 2.8\,{\mathrm{kHz}},
\end{cases}
\end{equation}
where radii are in km and frequencies in kHz.  Allowing for an estimated
uncertainty in the determination of $f_{\mathrm{peak}}$ and the maximum
deviation from this correlation for the different EoSs considered
in~\cite{2012PhRvD..86f3001B}, it is possible that a single observation of the
post-merger signal from a suriving PMNS could thus determine $R_{1.6}$ to an
accuracy of $100\text{--}200$\,m.  A relation between the dominant oscillation
frequency and neutron star radii has been confirmed
in~\cite{2013PhRvD..88d4026H}, and an attempt to infer the neutron star
compactness from a secondary peak is included in~\cite{2014arXiv1403.5672T}. 

A single observation of the post-merger  \gw{} signal from delayed or prompt
collapse can also constrain the threshold mass $M_{\mathrm{thresh}}$ for prompt
collapse. In the case of a delayed collapse observation, the total mass measured
from the inspiral signal provides a lower limit on $M_{\mathrm{thresh}}$, and an
upper limit can be inferred from the peak frequency $f_{\mathrm{peak}}$ and the
fact that the frequency increases with the mass of the remnant~\cite{2013PhRvL.111m1101B}. In the case of
unambiguous identification of a prompt collapse, the total mass measured from
the inspiral signal represents an upper limit on $M_{\mathrm{thresh}}$, leading
to constraints on the maximum mass of a non-rotating star in
isolation~\cite{2013PhRvL.111m1101B}.  Similarly, recent work has demonstrated
how two or more measurements of $f_{\mathrm{peak}}$ from systems with slightly
different masses may allow the determination of the maximum mass of cold,
non-rotating neutron stars to within $0.1\,\msun$ and the corresponding radius
to within a few percent~\cite{2014arXiv1403.5301B}. 

Finally, with the projected Einstein Telescope~\cite{2010CQGra..27h4007P} it may even be possible to use the post-merger signal to measure the
rest-frame source mass and luminosity distance of a BNS
system~\cite{2013arXiv1312.1862M}.  This measurement would break the
mass-redshift degeneracy present in observations of the inspiral phase and
permit the use of coalescing neutron stars as \emph{standard sirens} with \gw{}
observations alone~\cite{1986Natur.323..310S}.

Most detectability estimates for these systems in the literature generally find
that the post-merger signal may be detectable in aLIGO to distances of
$\sim\text{few}\text{--}20$\,Mpc, assuming an optimally oriented, overhead source
and that an optimal signal-to-noise ratio (SNR) of $\sim 5$ is sufficient for
detection~\footnote{We note that the simulations reported
in~\cite{2014arXiv1403.5672T} result in higher signal-to-noise ratios, increasing the
detection horizon to $20\text{--}40$\,Mpc, although these results do not seem
compatible with those reported in~\cite{2013PhRvD..88d4042R}.}.
%
%
However, the post-merger signal has only recently begun to be
described by analytical waveforms~\cite{2013PhRvD..88d4026H}.  Consequently, to search for and
characterize these signals, one must presently use more general
morphology-independent search techniques.  Here,  transient bursts of \gw{s} can
be identified in the detector output data as excess power localized in the
time-frequency domain (see
e.g.,~\cite{2001PhRvD..63d2003A,2004CQGra..21S1809C,2004CQGra..21S1819K,2010NJPh...12e3034S})
and the impinging \gw{} waveform can be reconstructed by considering the
coherent signal energy coincident in multiple
detectors~\cite{Klimenko:2005wa,Klimenko:2007hd,2008ApJ...678.1142S} or by
projecting the data onto bases formed from representative catalogues of
simulations of the un-modelled
signal~\cite{2009PhRvD..80j2004R,2012PhRvD..86d4023L}.  Additionally, in the
case of the high-frequency \gw{} burst following a binary neutron star
coalescence, it is reasonable to assume that the time of coalescence will be
known to high accuracy from the inspiral portion of the
signal~\cite{2005PhRvD..71h4008A,2011CQGra..28j5021F}, thus increasing the
detection confidence for the post-merger part.

It is the goal of this work to determine realistic estimates for the
detectability of the post-merger signal in the second generation of ground-based
\gw{} observatories using \cwb{}~\cite{Klimenko:2007hd}, which is the algorithm
used for unmodeled searches of gravitational wave
transients~\cite{2012PhRvD..85l2007A,Abadie:2012aa}, and simulated BNS merger
waveforms using a variety of EoSs.  We also introduce a novel and
computationally inexpensive algorithm for the analysis of the waveforms
reconstructed by CWB which allows accurate determination of the outcome of the
merger (delayed vs prompt collapse) and, where appropriate, a measurement of the
dominant post-merger oscillation frequency.

This paper is structured as follows:  in section~\ref{sec:algorithm} we describe
the data analysis algorithm and model selection procedure proposed for the
detection and characterisation of the post-merger \gw{} signal;
section~\ref{sec:experiment} describes the experimental setup of the study,
including a description of the LIGO data (\S~\ref{sec:data}) and the post-merger
waveform simulations (\S~\ref{sec:simulations}) used; in
section~\ref{sec:detectability} we describe the results of our analysis in terms
of the distance reach, expected detection rates and potential measurement
accuracy using the algorithm proposed in this work.   We conclude in
section~\ref{sec:conclusion} with a discussion of our findings and future
prospects.

\section{Analysis Algorithm}
\label{sec:algorithm}
In this section, we  describe the algorithm used to detect, classify and infer
the parameters of the putative post-merger signal, in two stages:
\begin{enumerate}
\item {\em Coherent Excess Power Detection}:  we use the Coherent WaveBurst
(CWB) algorithm~\cite{Klimenko:2007hd} to detect statistically significant high
frequency \gw{} signal power in the data stream around the time of a known BNS
coalescence. Once a signal is identified as significant with respect to the
noise via a constrained likelihood statistic, the CWB algorithm reconstructs the
detector responses using a coherent network analysis~\cite{Klimenko:2005wa}.  
\item {\em Signal Classification and Characterization}: Spectral analysis of
this reconstructed response from the first step is used to determine the outcome
of the merger (delayed collapse and a surviving PMNS vs prompt collapse to a
BH).  If the outcome is identified as a surviving PMNS, the dominant post-merger
frequency is recovered and used with equation~\ref{eq:fpeak_radius} to determine
the radius of a fiducial $1.6\msun$ neutron star as
in~\cite{2012PhRvD..86f3001B}.
\end{enumerate}

\subsection{Coherent WaveBurst}
\label{sec:cwb}
Searches for the inspiral \gw{} signal from coalescing binaries are typically
carried out using a matched-filtering technique and potentially large template
banks~\cite{2012PhRvD..85l2006A,2012PhRvD..85h2002A,Aasi:2013aa}.  The size and
composition of these template banks is defined by the details of the targeted
sources.  While analytical expressions are available and appropriate for the
inspiral portion of a BNS~\cite{lrr-2014-2}, phenomenological waveform families
including the merger and ring-down are needed for higher mass binary black hole
systems where the later part of the signal contributes significant
SNR~\cite{2007PhRvD..76j4049B,2011PhRvL.106x1101A}.  These families, however,
are not adequate for a post-merger search,  since they do not  yet include the
effects of the neutron star matter on the orbital evolution during the inspiral,
and they assume that the post-merger signal is the simple, quasi-normal mode
ringdown expected from a Kerr black hole (e.g.,~\cite{1989PhRvD..40.3194E}).
While in the past, binary neutron star simulations were performed with a simple
polytropic EoS and were focusing on the inspiral phase, most recent simulations
are including microphysical EoS and also focus on the long-term post-merger
evolution. However, the currently allowed sample of proposed EoS leads to a
variety of different outcomes. This motivates us to consider a hierachical
search approach, where the inspiral phase of the signal is detected via
matched-filtering to post-Newtonian analytical waveforms and then followed-up
using a morphology-independent analysis which identifies the post-merger signal.

The \cwb{} algorithm is designed to identify and reconstruct generic transients
in data collected from a network of interferometers. First, the data is
decomposed into {\em pixelated} maps, where each pixel represents the localized
energy of the data in a given time-frequency region.  Clusters of time-frequency
pixels across different interferometers' maps are marked as having significant
energy above the expected properties of the noise. Next, the analysis attempts
to match the expected response of a passing gravitational wave --- the two
independent polarizations denoted $h_+, h_{\times}$ --- in the network with a
maximum likelihood estimator:

\begin{equation}
\log L(h_+, h_{\times}) = \sum_{\Omega}
2\boldsymbol{x}\cdot\boldsymbol{\xi}-|\boldsymbol{\xi}|^2,
\end{equation}
where the boldfaced symbols imply a vector quantity formed from each member of
the network. The detector response
$\boldsymbol{\xi}=\boldsymbol{F_+}h_++\boldsymbol{F_{\times}}h_{\times}$
represents the inferred signal in the data $\boldsymbol{x}$, such that
$\boldsymbol{x}=\boldsymbol{n}+\boldsymbol{\xi}$ and $\boldsymbol{n}$ is the
intrinsic interferometer noise.  The dependence on source sky-location is
encoded in the geometrical antenna patterns $\boldsymbol{F_+},
\boldsymbol{F_{\times}}$, defined in~\cite{1998PhRvD..58f3001J}.  The two
polarizations of the GW signal are free parameters in the likelihood statistic
and $\Omega$ is the event's time-frequency area.

The likelihood for a time-frequency cluster is maximized over the source sky
location and the waveform is reconstructed as:

\begin{eqnarray}
h_+ = \frac{ \boldsymbol{F_+}\cdot{\boldsymbol{x}} }{ |\boldsymbol{F_+}|^2 } \\
h_{\times} = \frac{ \boldsymbol{F_{\times}}\cdot{\boldsymbol{x}} }{ |\boldsymbol{F_{\times}}|^2 }
\end{eqnarray}
The likelihood is an optimal statistic under the assumption that the detector
noise is stationary and Gaussian. In general, \gw{} detectors can suffer from
non-astrophysical, environmental, mechanical, or electrically induced
transients, referred to as \emph{glitches}. To mitigate the effects of these
glitches on the analysis and  reject false positives, several statistics
characterizing the consistency of the signal between interferometers, as well as
additional likelihood constraints (e.g.  imposing constraints on the
polarization content of the signal) have been developed and applied in previous
analyses~\cite{2009CQGra..26t4004P,Abadie:2012aa}.  None of these additional
constraints were utilized in this analysis as the detector data at high
frequencies is dominated by photon shot noise and is far less contaminated by
instrumental glitches than at lower frequencies.

Two statistics derived from the likelihood are used to identify and characterize
potential \gw{} events: the coherent network amplitude $\eta$, which is
proportional to the signal-to-noise ratio and is used to rank candidate events
and establish their significance, and the network correlation coefficient $cc$,
which is a measure of the degree of correlation between the detectors.  Both
statistics are described in detail in~\cite{s5_allsky}.  Small values of $cc \ll
1$ are typical for uncorrelated background events, while true \gw{} signals have
$cc$ close to unity.  A threshold of $cc = 0.5$ is used in the generation of
\cwb{} events in this analysis.  Determination of event significance using
$\eta$ and the \gw{} detection criterion is described in
sec.~\ref{sec:detection_criterion}.

\subsection{Post-merger Signal Classification \& Characterization}
\label{sec:classification}

\begin{figure*}
\includegraphics{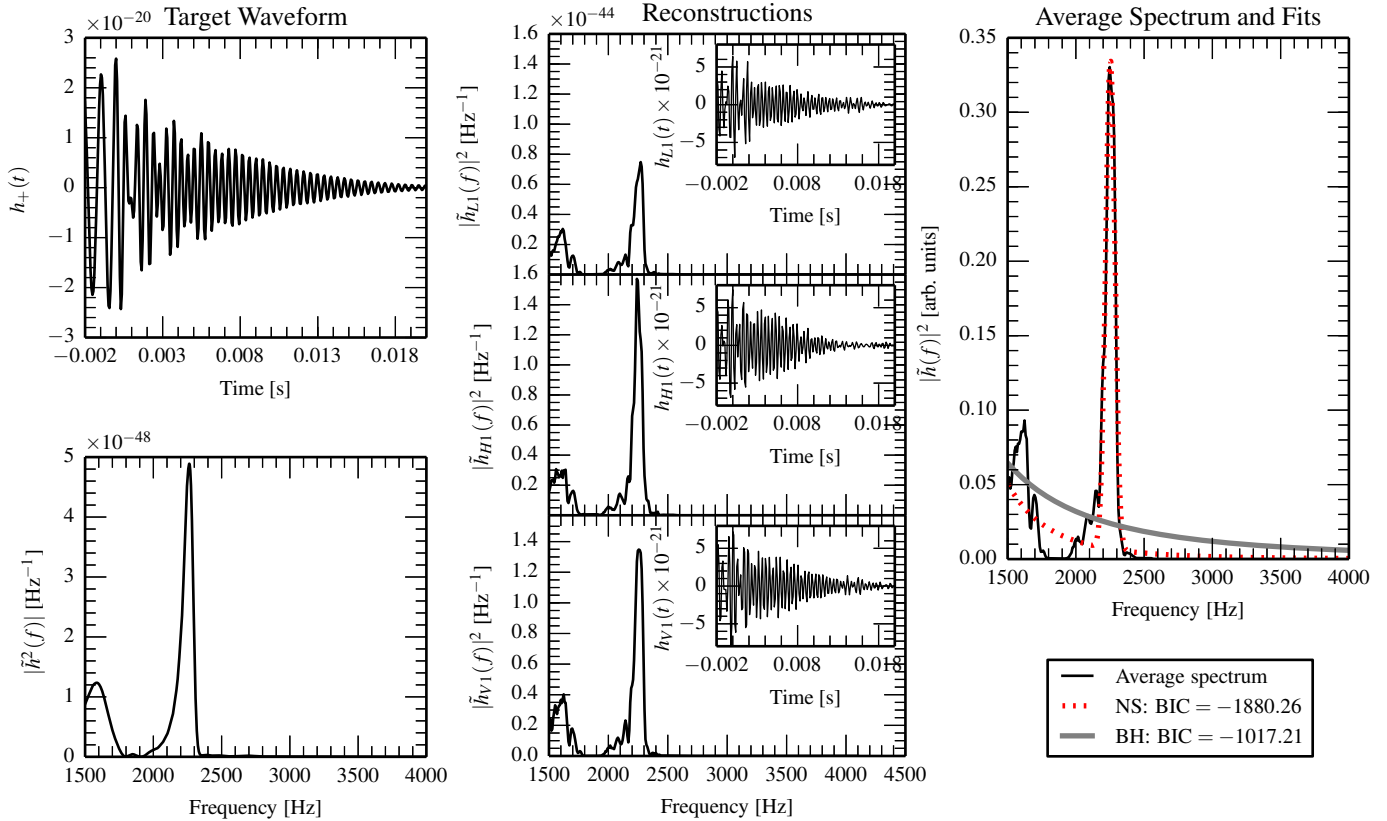}
\caption{Demonstration of signal characterization from the Shen EoS and
$1.35\text{--}1.35$\,$\msun$ system, which results in a surviving PMNS.
\emph{Left column}: the time-series and power spectral density of the `plus'
($+$) polarisation of the target waveform, for a source located 0.7\,Mpc from
the Earth.  A small distance is deliberately chosen to provide a high SNR signal
for demonstrative purposes.
\emph{Center column}: the power spectral densities and time series
(insets) of the detector responses reconstructed by the \cwb{} algorithm.  The
subscripts H1, L1 and V1 refer to simulated results from the LIGO detectors located
in Hanford and Livingston, and the Virgo detector, respectively.
\emph{Right column}: the SNR-weighted average reconstructed power spectral
density and fitted models.  The model for the delayed collapse scenario is
preferred in this instance, as indicated by the relative values of the Bayesian
Information Criterion (BIC), defined by equation~\ref{eq:bic}, for the delayed
and prompt collapse scenarios.\label{fig:pmns_example}}
\end{figure*}

\begin{figure*}
\includegraphics{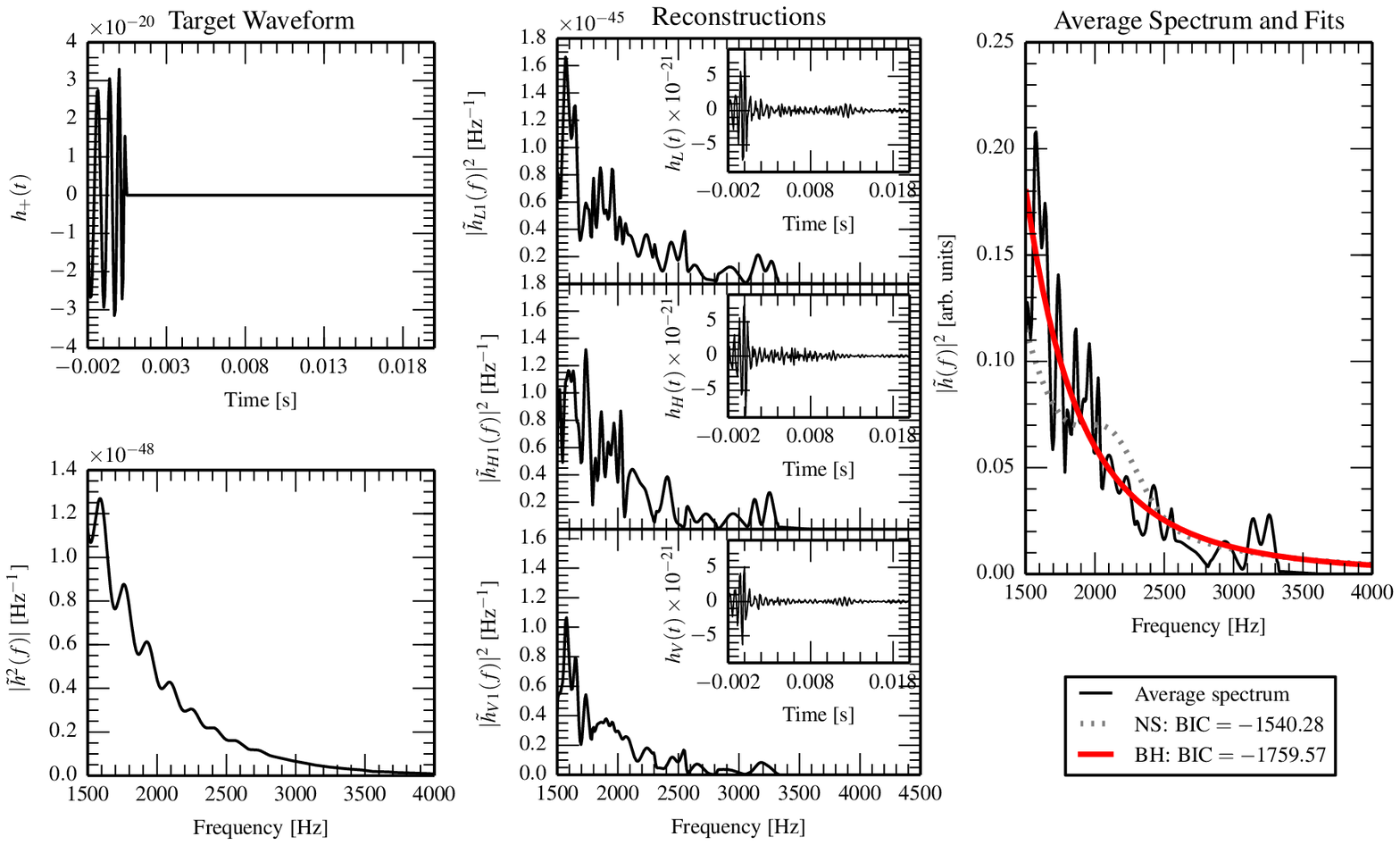}
\caption{Demonstration of signal characterization for the SFHo EoS and
$1.6\text{--}1.6$\,$\msun$ system, which results in prompt collapse to a BH.  BH
quasi-normal ringing is not included in the numerical approach used here but
lies at higher frequencies than are considered in this analysis.  
\emph{Left column}: the time-series and power spectral density of the `plus'
($+$) polarisation of the target waveform, for a source located 0.8\,Mpc from
the Earth.  A small distance is deliberately chosen to provide a high SNR signal
for demonstrative purposes.
\emph{Center column}: the power spectral densities and time series (insets) of
the detector responses reconstructed by the \cwb{} algorithm.  The
subscripts H1, L1 and V1 refer to simulated results from the LIGO detectors located
in Hanford and Livingston, and the Virgo detector, respectively.
\emph{Right column}: the SNR-weighted average reconstructed power spectral
density and fitted models.  The model for the delayed collapse scenario is
preferred in this instance, as indicated by the relative values of the BIC for
the delayed and prompt collapse scenarios.\label{fig:bh_example}}
\end{figure*}

A characteristic feature of the PMNS oscillation waveform is a distinct peak in
the power spectrum around $2\text{--}4$\,kHz with a bandwidth of several tens of
Hz.  This is in addition to the roughly power-law decay across frequency from
the late inspiral and merger, as well as one or more secondary oscillation
peaks.  An example
of a typical PMNS waveform may be found in figure~\ref{fig:pmns_example}. 
On the other hand, in the prompt collapse scenario, one still expects power across these
frequencies from the late inspiral and merger but any post-merger oscillation
comes from the stellar-mass black hole ring-down at $\gtrsim 6$\,kHz.
The waveform shown in figure~\ref{fig:bh_example} provides an example of the
signal expected from prompt collapse.

These features in the \gw{} signal spectrum may be identified in the waveforms
reconstructed by the \cwb{} algorithm, for candidate events that follow a
detected low-mass binary inspiral.  For this, we build an SNR-weighted average
power spectral density (PSD) of the reconstructed waveform in a network of
$N_{\textrm{det}}$ detectors as:
\begin{equation}\label{eq:average_PSD}
P_i = \frac{1}{N_{\textrm{det}}}\sum_{j=1}^{N_{\textrm{det}}}
\frac{\rho^{\mathrm{rec}}_j}  {\max_k (\rho^{\mathrm{rec}}_k)}P_{ij},
\end{equation}
where $i$ indexes the frequency bins  and $\rho^{\mathrm{rec}}_j$ is the SNR in
the $j$-th detector.  We model the PSD for the delayed collapse  as the sum of a
power law and a Gaussian:
\begin{equation}\label{eq:gauss_plus_powlaw}
S_{\textrm{NS}}(f) = A_0 \exp\left[ -\left( \frac{f-f^{'}_{\textrm{peak}}} {2\sigma}
\right)^2 \right] + A_1 \left(\frac{f}{f_{\textrm{low}}}\right)^\alpha,
\end{equation}
where $f_{\textrm{peak}}^{'}$ is an estimator for the true peak frequency
$f_{\mathrm{peak}}$ of the post-merger peak, and $\sigma$ is its characteristic
bandwidth.  $f_{\mathrm{low}}$ is the lower bound on the frequencies analysed,
$\alpha$ is the power law for the decay component of the signal and the terms
$A_0$ and $A_1$ set the amplitude scale of each component.

Since the post-merger signal is likely to be detectable only to ${\mathcal
O}(10)$\,Mpc we assume that the inspiral portion for the signal is detected with
high confidence (e.g., SNR $\sim 160$, at design
sensitivity~\cite{2013arXiv1304.0670L}).
Even at quite moderate SNRs (e.g., SNR $\sim 10$), the chirp mass ${\mathcal M}
= (m_1 m_2)^{3/5}(m_1 + m_2)^{-1/5}$ and symmetric mass ratio $\eta =
m_1m_2/(m_1+m_2)^2$ can be measured from the inspiral signal, with fractional
uncertainties as low as $\Delta {\mathcal M}/{\mathcal M} \lesssim 0.1\%$ and
$\Delta \eta / \eta \sim 1\text{--}10\%$~\cite{1993PhRvD..47.2198F,
1994PhRvD..49.2658C, 1996CQGra..13.1279J, 2005PhRvD..71h4008A,
2012PhRvD..85j4045V, 2013ApJ...766L..14H, 2013PhRvD..88f2001A}, which results in
a total mass uncertainty of a few percent.

The total mass of the system sets a lower bound on the expected
$f_{\mathrm{peak}}$.  Stars with a stiff equation of state are relatively
under-dense resulting in a low $f_{\mathrm{peak}}$.  The $f_{\mathrm{peak}}$ for
the stiffest equation of state (i.e., Shen) therefore represents a conservative
lower limit on the probable value for a given mass configuration and unknown
EoS.  We probe frequencies up to 400\,Hz below these values to account for
uncertainty in the mass measurement and in the EoS.  Lower bounds on the value
of $f_{\mathrm{peak}}$ used for the different mass configurations are given in
table~\ref{table:fpeak_bounds}.  Setting a lower frequency bound reduces the
chance that the Gaussian component of equation~\ref{eq:gauss_plus_powlaw} is
fitted to secondary, lower frequency peaks in the reconstructed spectrum and
significantly improves the robustness of the analysis.  We place an \emph{upper}
bound on $f_{\mathrm{peak}}$ at 4\,kHz; high enough to allow for the post-merger
ringing from softer (i.e., high frequency) EoS, such as APR, and low enough that
we expect no contribution from any black hole ringing, should the system undergo
gravitational collapse.

%
%
%
%
%
%
%
%
%

\begin{table}
\centering
\begin{tabular}{ l c c c }
\toprule
Total Mass $[\msun]$ & $f_{\mathrm{peak}}^{\mathrm{stiff}}$ [kHz] &
$\hat{f}_{\mathrm{peak}}^{\mathrm{min}}$ [kHz]\\ 
\cline{1-1}\cline{2-2}\cline{3-3}\cline{4-4}
2.7 (1.35, 1.35)    & 2.15 & 1.75 \\
3.2 (1.6, 1.6)      & 2.36 & 1.96 \\
3.3 (1.65, 1.65)    & 2.40 & 2.00 \\
3.8 (1.9, 1.9)      & 2.63 & 2.23 \\
\botrule
\end{tabular}
\caption{Estimates of the peak frequency of the post-merger \gw{} signal for the
stiffest EoS for the different masses considered in this analysis,
$f_{\mathrm{peak}}^{\mathrm{stiff}}$.  The analysis searches for spectral peaks
above $f_{\mathrm{peak}}^{\mathrm{min}}$.}
\label{table:fpeak_bounds}
\end{table}

In the case of prompt-collapse to a BH there will still be detectable signal
power from the late inspiral and merger.  Neglecting any contribution from the
BH ringdown, which is at significantly higher frequencies, we model the
reconstructed power spectrum as a power law,
\begin{equation}\label{eq:powlaw}
S_{\textrm{BH}}(f) = A_1 \left(\frac{f}{f_{\textrm{low}}}\right)^\alpha,
\end{equation}
where the terms are the same as those in equation~\ref{eq:gauss_plus_powlaw}. 

We select between these two models for the reconstructed PSD using the Bayesian
Information Criterion (BIC)~\cite{BIC_Schwarz}, defined as:
\begin{equation}\label{eq:bic}
\mathrm{BIC} = - 2 \ln {\mathcal L}_{\mathrm{max}} + k \ln n,
\end{equation}
where $n$ is the number of spectral bins analysed, $k$ is the number of free
parameters in the model and ${\mathcal L}_{\mathrm{max}}$ is the maximum
likelihood.  The BIC arises from approximating the relative Bayesian posterior
probabilities of models and provides a convenient measure of goodness-of-fit,
weighted by the parsimony of the model.  The model with the smallest value of
the BIC is preferred.  Assuming that the measurement errors are independent and
identically distributed according to a normal distribution, the BIC is, up to an
additive constant which is the same for all models:
\begin{equation}
\mathrm{BIC} = n \ln \chi^2_{\textrm{min}} + k \ln n,
\end{equation}
and
\begin{equation}
\chi^2_{\textrm{min}} = \frac{1}{n-1} \sum_{i=1}^n (P_i - S_i^*)^2,
\end{equation}
where $P_i$ and $S_i^*$ are the average power spectral density of the
reconstructed detector response and the value of the best-fitting model in the
$i^{\textrm{th}}$ frequency bin, respectively.  The best-fit model is found via
least-squares minimisation where the value of the center frequency of the
Gaussian component $f^{'}_{\mathrm{peak}}$ is constrained to lie above the
relevant value from table~\ref{table:fpeak_bounds} and the power law is
constrained such that $\alpha<0$.

\begin{figure}
\scalebox{0.7}{\includegraphics{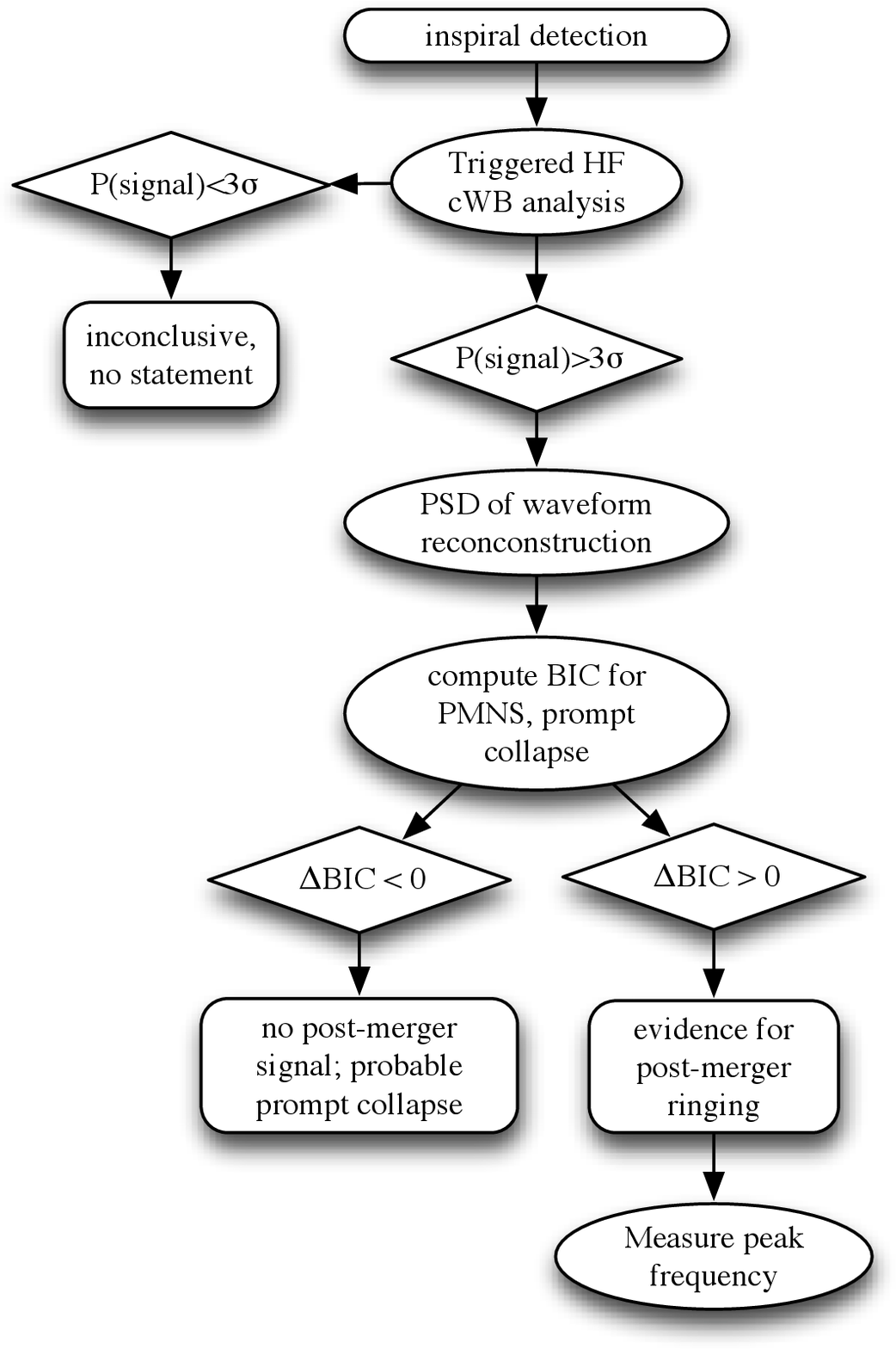}}
\caption{Proposed data analysis pipeline for the detection and characterization
of high-frequency \gw{}~signals following binary neutron star coalescence.  The
CWB un-modelled analysis algorithm is used to detect and reconstruct a
high-frequency component to the \gw{} signal temporally coincident with the
inspiral signal from BNS coalescence.  The Bayesian Information Criterion (BIC)
is then used to select between models for the frequency content for the
post-merger signals to determine whether the outcome of the BNS merger was
prompt collapse to a black hole or the formation of a post-merger neutron star
remnant.\label{fig:pipeline}}
\end{figure}

Figure~\ref{fig:pipeline} shows the workflow of this detection and
classification analysis pipeline.  The proposed procedure is:
\begin{enumerate}
\item We assume a robust detection of an inspiral signal from BNS is achieved
from a separate analysis, providing an estimate for the time of coalescence and
total mass of the system.
\item A high-frequency CWB analysis is performed in a small time window around
the time of coalescence of the BNS inspiral.  The CWB analysis is constrained 
 to $[1.5,~4]$\,kHz.
\item If CWB detects statistically significant excess power in a small time
window around the time of BNS coalescence, assume this is associated with the
coalescence and attempt to classify as follows.
\item Construct PSDs of detector reconstructions, $\{P_i\}_j$ and average
according to equation~\ref{eq:average_PSD} to obtain $\{P_i\}$.
\item Fit models described by
equations~\ref{eq:gauss_plus_powlaw},~\ref{eq:powlaw} to $\{P_i\}$ and compute
$\Delta \mathrm{BIC} = \mathrm{BIC}_{\mathrm{BH}} - \mathrm{BIC}_{\mathrm{NS}}$.
\item If $\Delta \mathrm{BIC} > 0$, the PMNS model is preferred and the
best-fitting value of $f^{'}_{\mathrm{peak}}$ provides our estimate of the peak
frequency of the post-merger oscillations.
\end{enumerate}

\section{Experimental Setup}
\label{sec:experiment}

The efficacy of the method outlined in section~\ref{sec:algorithm} is determined
via Monte Carlo simulations in which simulated post-merger signals are
superimposed on realistic detector data.  In this section, we describe the data and
waveform simulations used.

\subsection{GW Detector Data}\label{sec:data}
Data acquired by aLIGO and AdV is unlikely to be Gaussian or stationary and it
is helpful to demonstrate that our analysis method is robust to such features by
analysing realistic detector data.  To do so, we use a week of data recorded by
the initial generation instruments in 2007, recolored to the advanced detector
design sensitivities, following the procedure in~\cite{ninja2}.  

\begin{figure}[h!]
\includegraphics{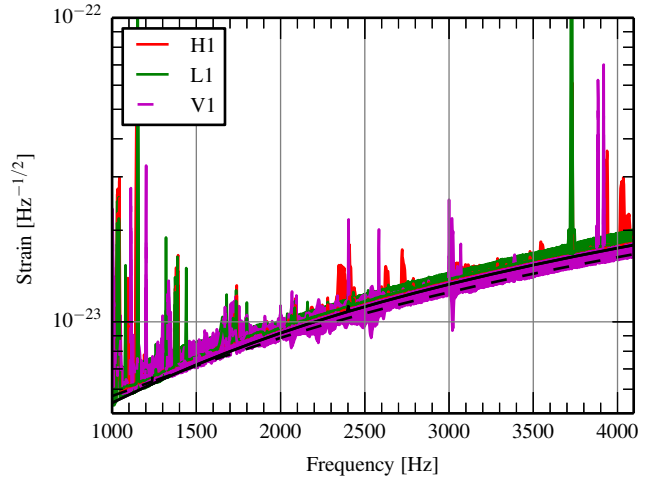}
\caption{The noise amplitude spectral density of the recolored data from the
LIGO (H1, L1) and Virgo (V1) detectors used in this analysis.  Shaded regions
(color online) indicate the $5^{\textrm{th}}$ (lower edge) and
$95^{\textrm{th}}$ (upper edge) percentiles of the variation in the noise floor
for the data used.  Black solid (dashed) curves indicate the design
sensitivities of the aLIGO (AdVirgo) detectors used for this
study~\cite{aLIGO_noise_curves,adV_noise_curves}.}
\label{fig:spectra}
\end{figure}

The LIGO data is recolored to have the Advanced LIGO design sensitivity given by
the zero-detuned, high-power noise curve~\cite{aLIGO_noise_curves}, while the
Virgo data is recolored to have the Advanced Virgo design sensitivity, given by
the dual recycled, 125 W, tuned signal recycling
configuration~\cite{adV_noise_curves}.  This choice of detectors and noise
curves loosely corresponds to a plausible \gw~detector network configuration for
c.2020~\cite{2013arXiv1304.0670L}.  Figure~\ref{fig:spectra} shows the noise
amplitude spectral densities (ASD) of the recoloured data for each detector,
where the colored regions indicate the variation between the $5^{\textrm{th}}$
and $95^{\textrm{th}}$ percentiles of the ASDs, measured over the analysed data.

The data from this period is not contiguous; the detectors were not always
operational and environmental artifacts and instrumental glitches affect the
quality of the data.  Such times are identified and removed from the analysis
following the procedures described
in~\cite{s5_highfreq,s5_allsky,s52vsr1_allsky}, leaving a total analysed time of
3.87 days.

%

\subsection{Binary Neutron Star Coalescence Simulations}
\label{sec:simulations}

\subsubsection{Merger Simulations}
\label{sec:simulation_details}

The waveforms used in our analysis are extracted from hydrodynamical simulations. These calculations are performed with a relativistic smooth particle hydrodynamics code, which employs the conformal flatness approximation for the solution of the Einstein field equations~\cite{1980grg..conf...23I,1996PhRvD..54.1317W}. Details on the numerical model can be found in~\cite{2002PhRvD..65j3005O,2007A&A...467..395O,2010PhRvD..82h4043B,2012PhRvD..86f3001B}. A comparison to fully relativistic grid-based simulations has revealed a quantitatively very good agreement~\cite{2012PhRvD..86f3001B,2013PhRvD..88d4026H,2014arXiv1403.5672T}. In comparison to~\cite{2012PhRvD..86f3001B} we implemented an improved version of the artifical viscosity scheme (see~\cite{1995JCoPh.121..357B,2007A&A...467..395O}) which reduces the artificial viscosity in dominantly rotational flows and causes less numerical damping of the fluid oscillations in the postmerger phase. The oscillation frequencies remain basically unchanged compared to the results presented in~\cite{2012PhRvD..86f3001B}, while the artificial damping of the GW amplitude is reduced.

The prime goal of this study is the extraction of equation-of-state properties
from the gravitational-wave detection of the neutron-star coalescence postmerger
phase. Since the properties of high-density matter are only incompletely known,
the numerical modeling relies on different theoretical descriptions of the
equation of state (see e.g.~\cite{2012ARNPS..62..485L} for a review). For this
work we employ a large variety of microphysical EoS models to ensure that the
full range of possible signatures is covered (see Tab.~\ref{tab:eos}). All
equations of state are compatible with the current lower limit on the maximum
mass of nonrotating neutron stars of about
2~$\msun$~\cite{2010Natur.467.1081D,Antoniadis26042013}. The considered
equations of state yield maximum gravitational masses of nonrotating neutron
stars from 2.02 to 2.79~$\msun$. Neutron star radii vary between 11.33~km and
14.75~km for 1.35~$\msun$ neutron stars and thus cover a significant part of the
range of typical radii constructed with various allowed EoS. Details on the
stellar properties for the specific models can be found in Tab.~\ref{tab:eos}
and in~\cite{2012PhRvD..86f3001B,2014arXiv1403.5301B}, which provide also the
mass-radius relations. All except for one equation of state (APR) take into account
the dependence on temperature and composition (electron/proton fraction). For
the APR model, which provides only the zero-temperature behavior, we employ an
approximate description of thermal effects (see e.g.~\cite{2010PhRvD..82h4043B},
which discusses also the reliability of the approximate treatment).

\begin{table}
\caption{\label{tab:eos} The nuclear equations of state used in this study. References are provided in the first column. Equations of state indicated by ``approx'' refer to models which rely on an approximate treatment of thermal effects, whereas ``full'' marks equations of state which provide the full temperature dependence. $M_{\mathrm{max}}$, $R_{\mathrm{max}}$, and $\rho_{\mathrm{c}}$ are the gravitational mass, circumferential radius, and central energy density of the maximum-mass Tolman-Oppenheimer-Volkoff configurations. We list $\rho_{\mathrm{c}}$ in units of the nuclear saturation density $\rho_0=2.7\times 10^{14}~\mathrm{g/cm^3}$. $R_{1.35}$ and $R_{1.6}$ are the circumferential radii of 1.35 and 1.6~$\msun$ neutron stars.}
\begin{ruledtabular}
\begin{tabular}{l|l|l|l|l|l}
EoS     & $M_{\mathrm{max}}$ & $R_{\mathrm{max}}$ &  $R_{1.35}$  &  $R_{1.6}$ & $\rho_{\mathrm{c}}/\rho_0$\\
        & $[\msun]$      & [km]               & [km]         &  [km]      &                           \\ \hline
APR~\cite{1998PhRvC..58.1804A} (approx)                     & 2.19 &  9.90  & 11.33 & 11.25 & 10.4  \\
NL3~\cite{1997PhRvC..55..540L,2010NuPhA.837..210H} (full)   & 2.79 & 13.43  & 14.75 & 14.81 & 5.6   \\
DD2~\cite{2010PhRvC..81a5803T,2010NuPhA.837..210H} (full)   & 2.42 & 11.90  & 13.21 & 13.26 & 7.2   \\
Shen~\cite{1998NuPhA.637..435S}  (full)                     & 2.22 & 13.12  & 14.56 & 14.46 & 6.7   \\
TM1~\cite{1994NuPhA.579..557S,2012ApJ...748...70H} (full)   & 2.21 & 12.57  & 14.49 & 14.36 & 6.7   \\
SFHX~\cite{2013ApJ...774...17S} (full)                      & 2.13 & 10.76  & 11.98 & 11.98 & 8.9   \\
SFHO~\cite{2013ApJ...774...17S}  (full)                     & 2.06 & 10.32  & 11.92 & 11.76 & 9.8   \\
TMA~\cite{1995NuPhA.588..357T,2012ApJ...748...70H} (full)   & 2.02 & 12.09  & 13.86 & 13.73 & 7.2   \\
\end{tabular}
\end{ruledtabular} 
\end{table}

The merger simulations start from quasi-equilibrium orbits a few revolutions before the coalescence. Initially, the temperature of the neutron stars is set to zero and the electron fraction is determined by neutrinoless beta-equilbrium. The intrinsic spin of neutron stars is assumed to be small compared to the orbital motion because the viscosity of neutron-star matter is not sufficient to yield tidally locked systems during the inspiral~\cite{1992ApJ...400..175B,1992ApJ...398..234K}. Hence, we adopt an irrotational velocity profile (see~\cite{2013arXiv1311.4443B} for an inclusion of spins in the case of an ideal gas equation of state).

Binary neutron star observations suggest (in accordance with population
synthesis studies) that symmetric systems with two stars of about $\sim
1.35\msun$ dominate the binary
population~\cite{2012ARNPS..62..485L,2012ApJ...759...52D}. Therefore, the
majority of waveforms used in this study are extracted from merger simulations
of equal-mass binaries with a total mass of 2.7~$\msun$, but we also explore cases with higher masses. It is worth noting that similar relations between the dominant GW oscillation frequency and fiducial neutron star radii exist also for other binary masses~\cite{Bauswein38Eos}. We leave the investigation of unequal-mass systems for the future, but note that the dominant oscillation frequency of the postmerger remnant resulting from asymmetric binaries is very close to the one from a symmetric merger of the same total mass (e.g.~\cite{2012PhRvD..86f3001B,2013PhRvD..88d4026H}).

For most investigated binary setups the merging results in the formation of a hot, massive, differentially rotating neutron star. The rapid differential rotation and thermal pressure support stabilize the remnant also in cases when the total binary mass exceeds the maximum mass of static non-rotating neutron stars. The collision induces strong oscillations, in particular, the quadrupolar fluid mode is strongly excited and generates the pronounced peak in the gravitational-wave spectrum (Fig.~\ref{fig:pmns_example}) (see~\cite{2011MNRAS.418..427S} for the identification of several oscillation modes in the merger remnant). After angular momentum redistribution and the extraction of energy and angular momentum by gravitational waves, the remnant possibly collapses to a black hole on a longer time, which typically exceeds the simulation time of about 20 ms after merging. The exact collapse time scale depends strongly on the total mass and also on other (partially not modelled) dissipative processes like magnetic fields, neutrino cooling and mass loss.

For sufficiently high total binary masses the remnant cannot be supported against the gravitational collapse and the merging leads to the direct formation of a black hole on a dynamical time scale. In our set of models the SFHo with 3.2~$\msun$ total binary mass represents such a case (see waveform in Fig.~\ref{fig:bh_example}). Note that our numerical approach does not allow to simulate the quasi-normal ringing of the BH. The oscillations of the BH occur at higher frequency than the $[1.5,~4.0]$\,kHz interval considered in this analysis and at smaller amplitude and, therefore, are unlikely to be confused with the signature of the NS postmerger remnants~\cite{2006PhRvD..73f4027S}. The threshold binary mass which results in the prompt collapse has been found to depend in a particular way on the equation of state and may yield information on the maximum mass of non-rotating neutron stars~\cite{2013PhRvL.111m1101B}. Therefore, we are also interested in distinguishing observationally the prompt collapse and the formation of a neutron star remnant. Note that the DD2 model with 3.2~$\msun$ total binary mass and the NL3 simulation with 3.8~$\msun$ total binary mass constitute models which are ``close'' to the prompt collapse because their binary masses are approximately 0.1~$\msun$ below the threshold. For both calculations the collapse still did not occur until the end of the simulation.

\subsubsection{Hybrid waveforms}
\label{sec:hybrid_waveforms}

The finite simulation time and the numerical damping of the postmerger
oscillations imply an underestimation of the actual GW amplitude. In an attempt
to accommondate this shortcoming of our approach we also include a set of hybrid
waveforms, constructed from a subset of the numerical waveforms described in the
previous section. We extend the numerical waveform with an analytically
prescribed waveform. The analytical part is described with a sinusoidal
waveform, which follows a prescribed frequency evolution and damping behavior.
The analytical model waveform is attached to the numerical waveform when the
numerical amplitue has decayed to one half of the initial postmerger GW
amplitude. This happens after several milliseconds when the remnant enters a
quasi-stationary phase. The initial frequency of the analytical waveform is
chosen to be the frequency of the GW signal at the matching point. We make
conservative assumptions about the further evolution of the frequency and the
damping timescale of the analytical model as explained below.

The damping of the postmerger oscillations and the evolution of the dominant oscillation frequency may be affected by different physical processes, such as gravitational wave emission, magnetic fields, neutrino heating and bulk viscosity (e.g.~\cite{2000ApJ...544..397S,2012PhRvD..86f4032P,2003pasb.conf..231R} and references therein). Here, we assume that the extraction of energy and angular momentum by gravitational waves is the dominant process responsible for the damping. Currently, there are no reliable estimates of the timescales of the other damping mechanisms, which is why we restrict ourselves to pure GW damping.

For cold, nonrotating NSs the damping timescale of the fundamental quadrupolar
fluid mode is known to depend on the star's mass and radius (see, e.g.
~\cite{1998MNRAS.299.1059A}). However, the postmerger remnant is a hypermassive
object rotating rapidly with strong differential rotation. For such a case,
there still exists no calculation of the actual damping timescales (see
\cite{FriedmanStergioulas} for the status on the subject).  

The damping timescales due to gravitational wave emission, assuming a quasi-stationary background, will be affected by a number of factors: a) rapid rotation, b) differential rotation, c) high mass, d) the equation of state, e) strong field gravity. In addition, if the background is evolving on a comparable timescale, then this will result in a time-dependent damping timescale. In the absence of a proper calculation that takes all of the above effects into account simultaneously, we are forced, at this point, to resort to some approximations in order to estimate upper and lower bounds for the expected damping timescale for each particular merger event we consider. Next, we give a detailed account of how we arrive at the particular upper and lower bounds used in the present work. We focus on the \textit{corotating}  $l=m=2$ $f-$mode, as this is the oscillation mode that is more likely to be excited during the merger of two neutron stars with a frequency of $\sim 2-3$ kHz. The corresponding counter-rotating mode will likely have a lower frequency in the inertial frame, as it is dragged towards corotation by rotation.

As an estimate for an \textit{upper bound} on the damping timescale, we apply the empirical formula from \cite{1998MNRAS.299.1059A}: 
\begin{equation}
{1 \over \tau_0[\rm s]}={\bar M^3 \over \bar R^4}\left [22.85-14.65\frac{\bar M}{\bar R} \right ], \label{eq:empirical}
\end{equation}
where $\tau_0$ is the damping timescale (in seconds) of an $l=2$ $f-$mode of a star of dimensionless mass $\bar M=M/1.4\,\msun$ and dimensionless radius
$\bar R = R/10\,{\rm km}$. Although the above formula was derived  for \textit{nonrotating} stars, we use it as an upper bound, since the actual damping timescale for rapidly rotating stars is shorter. Above, we
use the {\it mass of the remnant} (not of the individual components before merger) and we extract the equatorial radius of the remnant, neglecting its low-density envelope and consider the mass enclosed within this radius. For example, for the DD2 EoS we find an upper bound on the damping time scale of $\sim200$~ms for the remnant that results from the merger of two NSs with 1.35~$\msun$ each.
We note that the applicability of the above formula is limited only to remnants for which it still gives positive values for the damping timescale, i.e. to remnants for which ${\bar M \over \bar R} <1.56$.

As an estimate for a \textit{lower bound} on the damping timescale, we consider the following. In \cite{2011PhRvD..83f4031G,2013PhRvD..88d4052D} the damping timescale due
to gravitational wave emission of the $l=m=2$ $f-$mode in rapidly rotating
stars was studied, assuming uniform rotation and the Cowling approximation.  In particular,  \cite{2013PhRvD..88d4052D} used tabulated EoSs and estimated that the Cowling approximation overestimates the mode frequencies by up to $30\%$, while it underestimates damping timescales by up to a factor of three.
Nevertheless,    \cite{2013PhRvD..88d4052D} found an empirical relation between the damping timescale $\tau$ of a corotating $f-$mode in a uniformly rotating star and the corresponding damping timescale in a nonrotating model of \textit{the same central density}. This relation shows that  a star rotating at the mass-shedding limit will have a damping timescale which is $\sim 1/10$ of the corresponding nonrotating model with the same central density. We find that for remnants that are far from the threshold to prompt collapse, the central density of the remnant remains comparable, within a factor of two, to the central density of one of the binary component before merger. Therefore, one can relate the damping timescale of the rotating remnant to the damping timescale of a nonrotating model with mass equal to the mass of one of the binary components before merger, through the empirical relation found in     \cite{2013PhRvD..88d4052D}. We consider this as an \textit{approximate} lower bound, because the central density of the remnant is actually increasing somewhat, compared to the single star before merger, the actual damping timescale could be somewhat shorter, but at this level other uncertainties come into play and only a real calculation could give a precise result. 

For example, for a $ 1.35+1.35 \msun$ merger with the DD2 EoS, using Eq.~(\ref{eq:empirical}) for a nonrotating $ 1.35~\msun$ model, which has a radius of $\sim 13.2$km for this EoS, one obtains $\tau_0\sim280$~ms, and applying the empirical formula of  \cite{2013PhRvD..88d4052D}, this corresponds to  $\tau \sim 28$ms for a uniformly rotating star at the mass-shedding limit. For the same mass but with the APR EoS, we estimate a lower bound on the damping timescale of $\sim 18$ms; this is due to the smaller radius of $11.33$ km for this model. Note that the estimates of the GW emission timescale of $\sim30-50$ms in ~\cite{2005PhRvD..71h4021S,2005PhRvL..94t1101S} (which were based on the rate of angular momentum loss during simulations) fall within our estimated upper and lower bounds and are in fact
closer to our lower bound. 

The frequency of the corotating  $l=m=2$ $f-$mode levels off as the
mass-shedding limit is approached
(see~\cite{2010PhRvD..81h4055Z,2011PhRvD..83f4031G,2013PhRvD..88d4052D}) and so
does the  damping timescale~\cite{2013PhRvD..88d4052D}. In reality, the remnant
will rotate more rapidly than the mass-shedding limit for uniform rotation, but
it is evident from~\cite{2010PhRvD..81h4019K} that the frequency remains
practically constant, for reasonable values of the degree of differential
rotation. Therefore, to a first approximation we will neglect the effect of
differential rotation and assume that the reduction of the damping timescale by
a factor of $\sim 1/10$, as obtained at the mass-shedding limit for uniform
rotation, will also hold for hypermassive models with higher masses, but
comparable central density.    

   Finally, one should also consider the indirect effect of nonzero temperature on the damping timescales, through the corresponding increase in radius. For the temperatures occurring in the remnants in our simulations 
 this effect will be within the range of the upper and lower bounds considered above and so we do not treat it separately. This is based on the results of \cite{2011PhRvD..84d4017B}. 

We construct additional hybrid waveforms in which we allow for the remnant to become more compact during the evolution in order to test the sensitivity of our results to such an effect. In general, the $f-$mode frequency scales approximately 
as $\sqrt{M/R^3}$~\cite{2012PhRvL.108a1101B,2012PhRvD..86f3001B} and we consider the change of the frequency being mediated by a change of the remnant's radius. Here we assume that the loss of angular momentum by gravitational radiation is the dominant mechanism affecting the radius of the remnant while magnetic fields lead to a braking of the differential rotation on a timescale of $\sim 100$~ms~\cite{2012PhRvD..86f4032P}. The loss of angular momentum during the damping timescale of $\sim 200$~ms can be compared to the change of angular momentum in uniformly rotating neutron stars of constant rest mass. From this we obtain a rough estimate of the frequency change of about 5 per cent. We stress that mass loss counteracts the compactification of the remnant and that the frequency change is probably overestimated, in particular, for the hybrid waveforms with shorter damping timescales. We employ values of 5 and 0.0 per cent for the change of the dominant oscillation frequency per damping timescale. For shorter damping timescales a five per cent change represents a rather extreme case, which we choose by purpose to test the sensitivity of our method to such an extreme assumption.

Table~\ref{table:hybrid_characteristics} summarises the parameters of the
hybridized waveforms used in this study.

\begin{table}
\centering
\begin{tabular}{lccc}
\toprule
Hybrid & EoS & $\Delta f/f$\,[\%] & $\tau_0$\,[ms] \\
\colrule
hlAPR$^{\dagger}$ & APR & 0.00 & 180 \\
hlAPR$^{*}$ & APR & 0.05 & 200 \\
hsAPR$^{\dagger}$ & APR & 0.00 & 18 \\
hsAPR$^{*}$ & APR & 0.05 & 18 \\
hlDD2$^{\dagger}$ & DD2 & 0.00 & 200 \\
hlDD2$^{*}$ & DD2 & 0.05 & 200 \\
hsDD2$^{\dagger}$ & DD2 & 0.00 & 28 \\
hsDD2$^{*}$ & DD2 & 0.05 & 28 \\
\botrule
\end{tabular}
\caption{Characteristics of the hybridized waveforms used in this study.
Daggers ($\dagger$) and asterixes ($*$) indicate whether the analytic part of
the waveform is a stationary (in frequency) ring-down or a decaying chirp, where
the frequency increases by the percentage shown in the $\Delta f/f$ column.  The
$\tau_0$ column indicates the e-folding time for the decay of the analytic
signal.}
\label{table:hybrid_characteristics}
\end{table}

\subsubsection{\gw~Signal Simulations}
\label{sec:injections}
\begin{figure*}
\scalebox{0.9}{\includegraphics{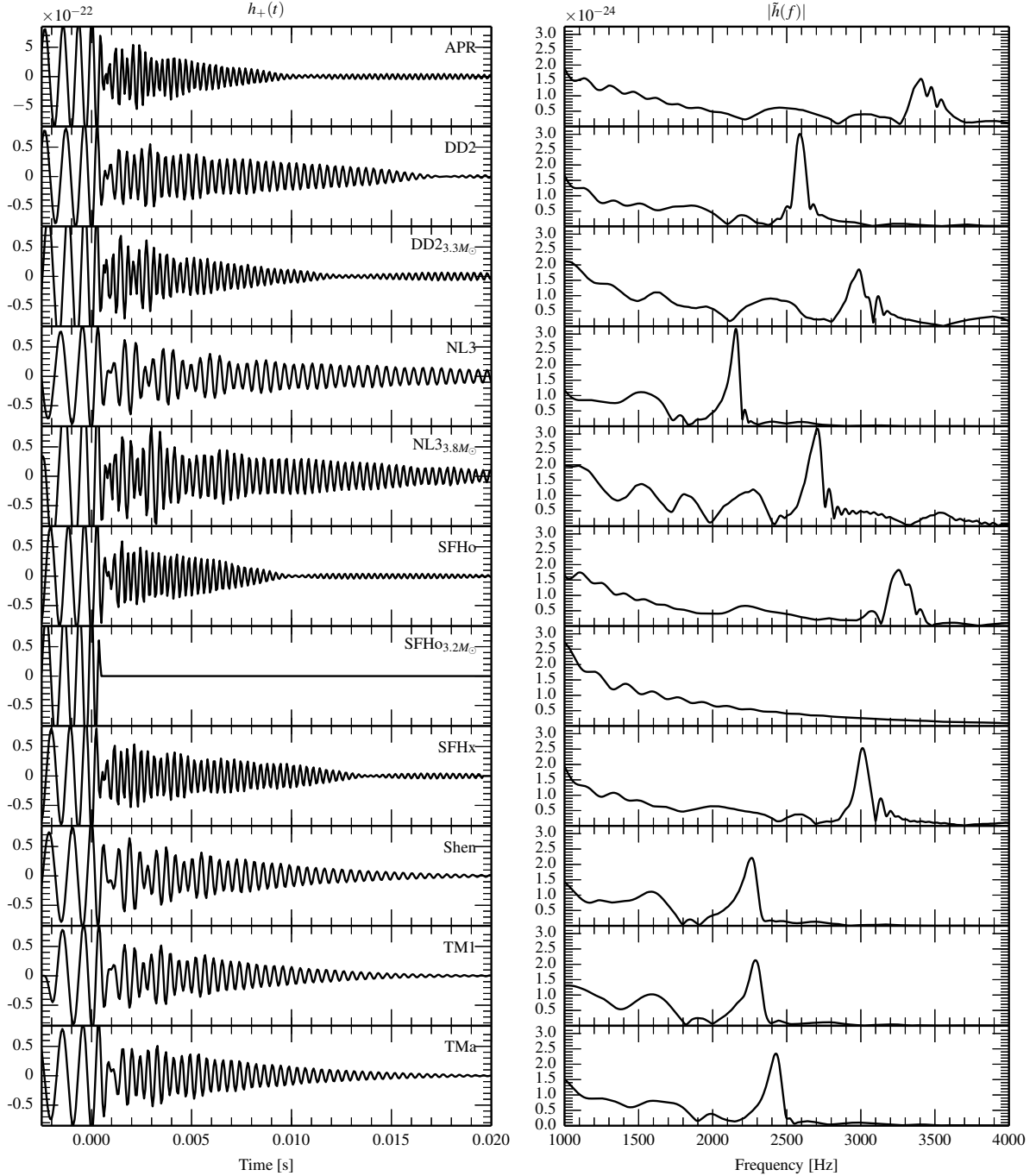}}
\caption{The catalogue of waveforms used in this study.  \emph{Left column}: The
time series of the plus polarization of the gravitational waves for a source at
20\,Mpc.  \emph{Right column}: The amplitude spectral density of the
characteristic strain (solid line) for an optimally located and oriented source
and the aLIGO design sensitivity (dashed line).}
\label{fig:catalogue}
\end{figure*}

\begin{figure*}
\scalebox{0.9}{\includegraphics{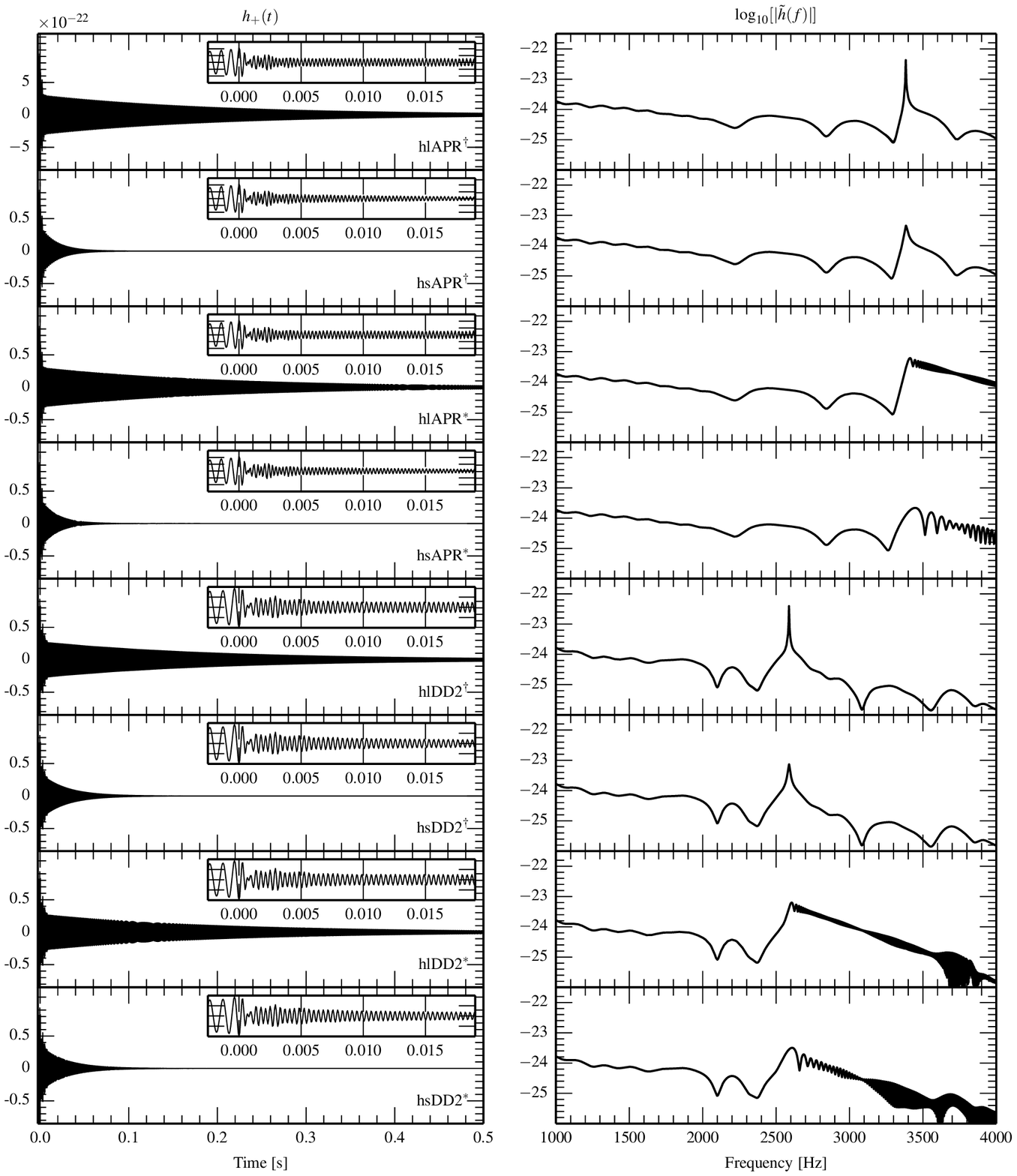}}
\caption{The hybridized waveforms used in this study.  \emph{Left column}: The
time series of the plus polarization of the gravitational waves for a source at
20\,Mpc.  Insets show the early evolution of the signal for comparison with the
original waveforms shown in figure~\ref{fig:catalogue}.
\emph{Right column}: The amplitude spectral density of the
characteristic strain (solid line) of the hybridized waveforms for an optimally
located and oriented source and the aLIGO design
sensitivity (dashed line).} \label{fig:hybrid_catalogue} 
\end{figure*}

The \gw~polarizations $h_+$ and $h_{\times}$ are computed from  the second time
derivative of the quadrupole moment of the source $\ddot{\mathbf{I}}$, which is
obtained from the numerical simulations.  The quantity $h_+ - ih_{\times}$ can
be decomposed into modes with spin weighted spherical harmonics
$\leftidx{^{s}}Y_{\ell m}(\theta,\phi)$ of weight -2:
\begin{equation}\label{eq:waveform_construction}
h_+ - ih_{\times} = \frac{1}{D} \sum_{\ell=2}^{\infty} \sum_{m=-\ell}^{\ell}
\leftidx{^{-2}}\, Y_{\ell m}(\theta,\phi)\,H_{\ell m}(t).
\end{equation}
The expansion parameters $H_{\ell m}(t)$ are complex functions of the retarded
source time $t$.  The $H_{2m}$, where $\ell=2$ is the quadrupole mode, may be expressed in terms
of the second time derivatives of the Cartesian components of the mass
quadrupole moment $\ddot{I}$ as,
\begin{eqnarray}\label{eq:expansion_params}
H_{20} & = & \sqrt{\frac{32\pi}{15}} \frac{G}{c^4} \left[ \ddot{I}_{zz} -
\frac{1}{2} \left(\ddot{I}_{xx} + \ddot{I}_{yy}\right) \right] \\
H_{2\pm1} & = & \sqrt{\frac{16\pi}{5}} \frac{G}{c^4} \left( \mp \ddot{I}_{xz} +
i \ddot{I}_{yz}\right) \\
H_{2\pm2} & = & \sqrt{\frac{4\pi}{5}}\frac{G}{c^4} \left( \ddot{I}_{xx} -
\ddot{I}_{yy} \mp 2i \ddot{I}_{xy}\right).
\end{eqnarray}
Figures~\ref{fig:catalogue} and~\ref{fig:hybrid_catalogue} show the catalogue of
waveforms used in this study, assuming a distance of 20\,Mpc and optimal source
sky-location and orientation.  We discuss the characteristics of these waveforms
in section~\ref{sec:characteristics}.

The polarisations $h_+$ and $h_{\times}$ constructed from the quadrupole moments
from each simulation were superimposed on the recolored data streams from each
detector after the appropriate projection onto the sky for a given sky
location, inclination and polarization:
\begin{equation}\label{eq:det_response}
h(t) = F_+({\mathbf \Omega},\psi) h_+(t) + F_{\times}({\mathbf \Omega},\psi)
h_{\times}(t),
\end{equation}
where ${\mathbf \Omega}$ is the sky-location, $\psi$ is the \gw~polarization
angle and $F_+$ and $F_{\times}$ are the detector antenna responses, defined
in~\cite{1998PhRvD..58f3001J}.  The inclination dependence enters through the
spherical harmonics in equation~\ref{eq:waveform_construction}.   Note that, the
amplitude of the waveforms is scaled up by $40\%$ before being injected to
account for the amplitude underestimate from extraction in the quadrupole
approximation~\cite{2005PhRvD..71h4021S,2012PhRvD..86f3001B}.

We refer to these signal simulations as \emph{injections}.  Injections are added
approximately every 60 seconds with a uniform random offset within a 10 second
window. This placed all injections far enough apart that the whitening and noise
estimation procedures, which use data surrounding each injection, is never
affected by a neighboring injection.  The injections were distributed
isotropically over the sky with randomized source inclinations (uniform in $\cos
\iota$) and polarizations (uniform in polarization angle $\Psi$).  Thus, all
results quoted in this study are averaged over sky location and source
orientation.  The distances of the injections are distributed uniformly in
$[0.5,~8]$\,Mpc for the purely numerical waveforms and in $[0.5,15]$\,Mpc for
the hybridized waveforms.  These distributions are chosen to cover the range of
detection scenarios and are not intended to correspond to astrophysical
scenarios.  This procedure was repeated ten times resulting in a population of
approximately 53000 injections for each waveform type.  The results of the
injections are then binned in distance and used to estimate detection
probability as a function of distance as described in \S~\ref{sec:results}.

We conduct separate simulation campaigns for each of the waveforms in our
catalogue (i.e., figure~\ref{fig:catalogue}, table~\ref{table:characteristics})
and characterise the results for each waveform separately.  

\subsection{Background Estimation \& Detection Criterion}
\label{sec:detection_criterion}

Following the approach used in recent searches for \gw{}
transients~\cite{2012PhRvD..85l2007A,Abadie:2012aa} and as described in
section~\ref{sec:cwb}, triggers arising from the CWB analysis are ranked by
their coherent network amplitude, $\eta$.  The statistical significance of CWB
triggers is determined from the distribution $\eta$ in the absence of
\gw~signals, the background distribution, which  is estimated by time-shifting
individual detector data streams relative to each other by an amount greater
than the light travel time between each detector.  This ensures no \gw{} signal
is present in the background data set, and also provides a convenient means to
increase the background statistics of a relatively short data segment.  Fifty
such time shifts in increments of 1 second are applied in this study.  This
background analysis assigns a false alarm rate (FAR) to the CWB triggers.  This
rate is then interpreted as a $p$-value by assuming some observation time.  For
the purposes of this study, we assume an observation time of $T_{{\mathrm
obs}}=100$\,ms.  This represents a conservative estimate of the time of
coalescence measured from the inspiral signal.

We place a threshold on the statistical significance required for
detection of $3\text{--}\sigma$.  The network configuration chosen for this
analysis corresponds to an expected BNS detection rate of ${\mathcal O}(100)$
events / year.  For a $3\text{--}\sigma$ significance then, with 100 trials,
triggers with $p$-value $p<10^{-5}$ are regarded as \gw~detection candidates.
Finally, the assumed observation time of $T_{{\mathrm obs}}=100$\,ms results in
a FAR threshold of $10^{-4}$\,Hz.

\section{Detectability Study}
\label{sec:detectability}
We now discuss the prospects for the detection and measurement of the
post-merger \gw{} signal with the algorithm described in
section~\ref{sec:algorithm}.

\subsection{Waveform Characteristics \& Expected Detectability}
\label{sec:characteristics}

We begin by considering the expected detectability of the post-merger signal
with an optimal matched-filter, While matched-filtering may not be realistic,
due to the scarcity of templates and  high computational costs, it provides an
estimate for the best-case sensitivity to these systems in stationary, Gaussian
data.

If the form of the expected \gw{} signal in the detector is known \emph{a priori},
the optimal detection statistic is the matched-filter signal-to-noise ratio
(SNR) $\rho$:
\begin{equation}
\rho^2 = 4\Re \int^{f_{\mathrm{upp}}}_{f_{\mathrm{low}}}
\frac{\tilde{d}(f)\tilde{h}^*(f)}{S(f)}\diff f,
\end{equation}
where $\tilde{h}(f)$ is a template for the expected \gw{} signal, $\tilde{d}(f)$
is the Fourier transform of the data, $S(f)$ is the  one-sided noise power
spectral density and $f_{\mathrm{low}}$ and $f_{\mathrm{upp}}$ are lower and
upper bounds on the searched frequency range~\cite{lrr-2012-4}. 

Under the assumption of Gaussian noise and in the absence of a signal, $\rho^2$
follows a central $\chi^2$-distribution with $k=2$ degrees of freedom.  The SNR
threshold $\rho_{\mathrm{thresh}}$ which corresponds to a false alarm
probability of $10^{-5}$ is found from the survival function of the SNR
distribution in Gaussian noise, evaluated at the chosen false alarm probability:
\begin{equation}\label{eq:fap}
\mathrm{FAP} = 1 - P_{\chi^2}(\rho^2 \leq \rho^2_{\mathrm{thresh}} | k=2)^{N_t}.
\end{equation}
In this equation, $N_t$ is a trials factor introduced by searching over a
template bank.  
For the most optimistic estimate, we assume the signal is known exactly and only
a single waveform template is required, so that $N_t=1~$\footnote{Note that this
also implies that the sky-location and time of the signal are known.}.  For
FAP=$10^{-5}$, eq.~\ref{eq:fap} yields $\rho_{\mathrm{thresh}}=4.8$.
%
Table~\ref{table:characteristics} lists the SNR for the waveform of this study,
evaluated at $20$\,Mpc.  Two SNRs are reported: SNR$_{\mathrm{full}}$; the SNR
evaluated over the full frequency range of $[1.5,~4]$\,kHz and which is used to
determine the detectability of the signal and SNR$_{\mathrm{peak}}$ which is the
SNR evaluated over a narrow frequency range around the dominant high-frequency
peak and indicates the relative strength of the post-merger oscillation as
compared with the full late-inspiral, merger and post-merger signal.   

The distance-reach of a search is often characterized by its \emph{horizon
distance} $D_{\mathrm{h}}$, the distance at which an optimally-oriented
 source yields an SNR at least as large as the detection threshold.
Since SNR
scales inversely with distance, the horizon distance is obtained by rescaling
the fiducial 20\,Mpc to that distance which yields SNR$_{\mathrm{full}} \approx
5$. 

Following~\cite{1993PhRvD..47.2198F}, we define the effective range ${\mathcal
R}_{\mathrm{Opt}}$ of this hypothetical, optimal search as the radius enclosing
a spherical volume $V$ such that the rate of detections from a homogenous,
isotropic distribution of sources with rate density $\dot{\mathcal{N}}$ is
$\dot{\mathcal{N}}V$.  For an elliptically polarized source~\cite{SuttonBursts} 
the effective range ${\mathcal R}_{\mathrm{Opt}}
\approx D_{\mathrm{h}}/2.26$, where the factor 2.26 accounts for the average
over all sky-locations and orientations.  Table~\ref{table:characteristics}
lists the optimal effective range ${\mathcal R}_{\mathrm{Opt}}$ for each waveform in the
catalogue, 
calculated from
the noise PSD of a single aLIGO instrument at design sensitivity.  An optimal
search with $X$ detectors with comparable sensitivity will be a factor
$\sqrt{X}$ more sensitive than the single detector search~\cite{ratesPaper}.  In
table~\ref{table:characteristics}, we also report the expected effective range for an
optimal search in Gaussian noise assuming a network of $3$ instruments with the
aLIGO design sensitivity.  Figure~\ref{fig:sens_dist} summarises the
theoretically-achievable effective range for each waveform and compares this
with the result from the CWB Monte-Carlo analysis reported in the next section.

The expected detection rate $\dot{N}_{\mathrm{det}}^{\mathrm{Opt}}$ is obtained
by considering the number of Milky Way Equivalent Galaxies (MWEGs) within the
effective range~\footnote{Found from figure~1 of~\cite{ratesPaper}} and the
estimated BNS coalescence rate.  Assuming the coalescence rate $\dot{\mathcal
N}=100$\,MWEG$^{-1}$\,Myr$^{-1}$~\cite{ratesPaper}, we find
$\dot{N}_{\mathrm{det}}^{\mathrm{Opt}} \sim 0.01\text{--}0.1$\,year$^{-1}$, in
reasonable agreement with previous estimates 
~\cite{2005PhRvL..94t1101S, 2007PhRvL..99l1102O, PhysRevD.78.084033,
2011PhRvD..83d4014G, 2012PhRvD..86f3001B, 2013PhRvD..88d4026H,
2014arXiv1403.5672T}.

Finally, table~\ref{table:characteristics} reports the energy emitted in \gw{s},
E$_{\mathrm{GW}}$ and the peak post-merger frequency, where appropriate.  The
energy $E_{\mathrm{GW}}$ was calculated from a numerical integration of:
\begin{eqnarray}\label{eq:energy}
E_{\mathrm{GW}} & = &\frac{\pi c^3}{4G}D^2 \int_{-1}^1 \diff(\cos\iota)
\int_0^{2\pi}\diff \lambda \\ \nonumber 
& \times & \int_{-\infty}^{\infty}\diff f
\left[\frac{(1+\cos^2 \iota)^2}{4} + \cos^2 \iota\right] f^2|\tilde{h}(f)|^2
\\\nonumber
& = & \frac{8\pi^2 c^3}{5G}D^2 \int_{-\infty}^{\infty}\diff f
f^2|\tilde{h}(f)|^2.
\end{eqnarray}

\begin{table*}
\begin{ruledtabular}
\centering
\begin{tabular}{  l   c   c   c   c   c   c  c}

Waveform & SNR$_{\textrm{full}}$ & SNR$_{\textrm{peak}}$ &${\mathcal R}_{\textrm{Opt}}$ [Mpc] &$\dot{N}^{\textrm{Opt}}_{\textrm{det}}$ [year$^{-1}$]&$E_{\mathrm{GW}}$ [$M_{\odot}$] & $E_{\mathrm{GW}}^{\mathrm{peak}}$ [$M_{\odot}$] &$f_{\textrm{peak}}$ [Hz]\\ 
\colrule
$\textrm{APR}$ & 4.07 & 1.66 & 13.00 & 0.04 &0.09 & 0.05 & 3405.40\\ 
$\textrm{DD2}$ & 4.19 & 3.13 & 13.38 & 0.04 &0.07 & 0.06 & 2588.60\\ 
$\textrm{DD2}_{3.3M_{\odot}}$ & 4.69 & 2.00 & 14.98 & 0.06 &0.09 & 0.04 & 2987.00\\ 
$\textrm{NL3}$ & 4.58 & 3.34 & 14.64 & 0.05 &0.04 & 0.03 & 2156.80\\ 
$\textrm{NL3}_{3.8M_{\odot}}$ & 5.89 & 3.46 & 18.82 & 0.12 &0.14 & 0.08 & 2706.60\\ 
$\textrm{SFHo}$ & 3.82 & 2.06 & 12.20 & 0.03 &0.08 & 0.06 & 3255.20\\ 
$\textrm{SFHo}_{3.2M_{\odot}}$ & 4.28 & - & 13.65 & 0.04 & 0.04 & - & -\\ 
$\textrm{SFHx}$ & 3.98 & 2.44 & 12.72 & 0.04 &0.09 & 0.06 & 3011.40\\ 
$\textrm{Shen}$ & 4.35 & 2.96 & 13.89 & 0.05 &0.04 & 0.03 & 2263.20\\ 
$\textrm{TM1}$ & 4.05 & 2.73 & 12.94 & 0.04 &0.04 & 0.03 & 2288.60\\ 
$\textrm{TMa}$ & 4.03 & 2.84 & 12.86 & 0.04 &0.05 & 0.04 & 2426.80\\ 
$\textrm{hlAPR}^{\dagger}$ & 7.67 & 5.54 & 24.51 & 0.25 &0.76 & 0.49 & 3383.40\\ 
$\textrm{hsAPR}^{\dagger}$ & 4.43 & 2.00 & 14.15 & 0.05 &0.14 & 0.06 & 3384.20\\ 
$\textrm{hlAPR}^{*}$ & 7.41 & 4.05 & 23.66 & 0.23 &0.82 & 0.27 & 3412.60\\ 
$\textrm{hsAPR}^{*}$ & 4.39 & 2.23 & 14.02 & 0.05 &0.14 & 0.09 & 3447.20\\ 
$\textrm{hlDD2}^{\dagger}$ & 8.49 & 6.74 & 27.12 & 0.30 &0.38 & 0.26 & 2587.80\\ 
$\textrm{hsDD2}^{\dagger}$ & 4.83 & 3.32 & 15.42 & 0.06 &0.10 & 0.06 & 2588.00\\ 
$\textrm{hlDD2}^{*}$ & 8.21 & 6.11 & 26.23 & 0.28 &0.41 & 0.22 & 2606.00\\ 
$\textrm{hsDD2}^{*}$ & 4.76 & 3.28 & 15.20 & 0.06 &0.11 & 0.06 & 2609.00\\ 
\end{tabular}

\caption{Characteristics of the waveforms used in this study.  Unless otherwise
indicated in the subscript, the total mass is $2.7\msun$ and all systems have a
symmetric mass configuration (see \S~\ref{sec:simulation_details}) .  Signal to
noise ratios are evaluated for an optimally oriented source at 20\,Mpc.  Fields
denoted ``full'' refer to quantities evaluated over the full frequency range
used for detection $[1.5,~4.0]$\,kHz.  Fields denoted ``peak'' refer to those
quantities evaluated in a narrow range around the dominant high-frequency
spectral peak (a $2\sigma$ width around the best fitting Gaussian).   ${\mathcal
R}_{\textrm{Opt}}$ is the effective range of an optimal matched-filter search,
assuming a $3\text{--}\sigma$ statistical significance and $\mathcal{O}(100)$
BNS-inspiral triggers; $\dot{N}_{\textrm{det}}^{\textrm{Opt}}$ is the expected
number of post-merger signal detections for a search with this effective range.
The energy, $E_{\mathrm{GW}}$ is the energy carried by the \gw{} signal assuming
elliptical polarization and computed according to equation~\ref{eq:energy}.
$f_{\mathrm{peak}}$ is the frequency of the highest peak in the signal power
spectrum.  Note that the $3.2\,\msun$ SFHo waveform exhibits prompt collapse so
the post-merger characteristics are undefined here.
\label{table:characteristics}}
\end{ruledtabular}
\end{table*}


\subsection{Monte-Carlo Study}
\label{sec:results}

\subsubsection{Detectability}
\begin{figure}
\scalebox{1}{\includegraphics{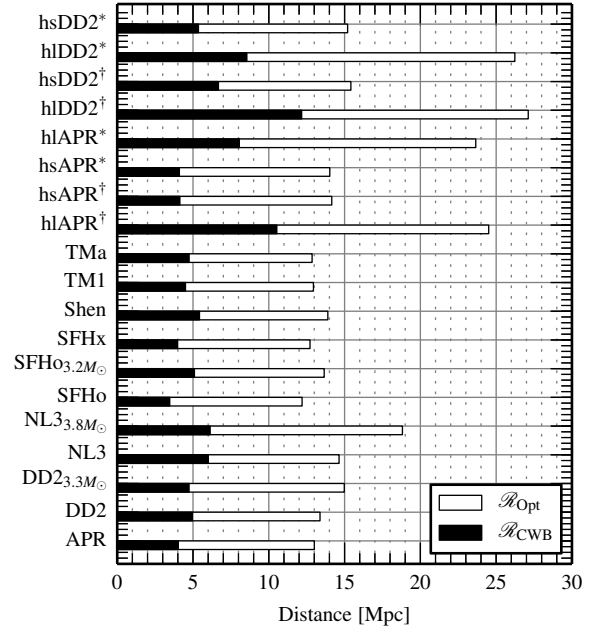}}
\caption{The effective ranges for each waveform for an idealized optimal
matched-filter analysis strategy (${\mathcal R}_{\mathrm{Opt}}$) and the \cwb 
Monte-Carlo study (${\mathcal R}_{\mathrm{CWB}}$).  Both ranges are evaluated at
assuming a false alarm probability of $10^{-5}$.  Differences in the ranges for
each waveform are consistent with the difference in the optimal
SNR.\label{fig:sens_dist}} 
\end{figure}

For a more realistic estimate, we performed a Monte-Carlo study with realistic data and the \cwb{}
algorithm, as presented in section~\ref{sec:algorithm}.
We define the analysis effective range ${\mathcal R}_{\mathrm{CWB}}$ 
as the radius of a volume $V$ such that
the rate of detections is $\dot{N}V$, where $V$ is:
\begin{equation}
V =  \int_0^{\infty} \diff r~4\pi r^2 \epsilon(r)
\end{equation}
and $\epsilon(r)$ is the probability of detecting the post-merger signal at
distance $r$, averaged over sky-location and orientation.  $\epsilon(r)$ is
referred to as the {\em efficiency} of the search, and is determined by
binning the injections in distance and counting the number of found injections
$k$ out of $N$ trials.  Assuming a uniform prior on the efficiency, the
posterior probability density distribution for $\epsilon$ is,
\begin{equation}
P(\epsilon|k,N) = \frac{\left(N+1\right)!}{\left(N-k\right)!k!}
\epsilon^k\left(1-\epsilon\right)^{N-k}.
\end{equation}
We estimate the efficiency from its expectation value, given $k$ detections in
$N$ simulations at distance $r$,
\begin{eqnarray}\label{eq:efficiency}
\langle \epsilon \rangle & = & \int \epsilon P(\epsilon|k,N)\diff \epsilon \\
& = &  \frac{k+1}{N+2}.
\end{eqnarray}
The effective range is the radius of a sphere of volume $V$:
\begin{equation}
\mathcal{R}_{\mathrm{CWB}} = \left[ 3 \int_0^{\infty} \diff r\, r^2\,
\epsilon(r)\right]^{1/3} .
\end{equation}
We report the values for $\mathcal{R}_{\mathrm{CWB}}$ for each waveform in the
catalogue in figure~\ref{fig:sens_dist} and in table~\ref{table:results}.  We
find that the range of the CWB analysis is approximately $60\text{--}70\%$
smaller than 
an optimal search with perfect
knowledge of the waveform in stationary, Gaussian noise.
This also implies a reduction of  the expected detection rate, with
$\dot{N}_{\mathrm{det}}^{\mathrm{CWB}} \sim 10^{-3}\text{--}0.1$ events/year, as
listed in table~\ref{table:results}.

\begin{table*}
\begin{ruledtabular}

\begin{tabular}{  l  c  c  c  c  c }
Waveform & ${\mathcal R}_{\mathrm{CWB}}$ [Mpc] & $\dot{N}_{\mathrm{det}}\times
10^{-2}$ [year$^{-1}$] & Classification Accuracy & $\tilde{\delta
f}_{\mathrm{peak}}$ [Hz] & $\tilde{\delta R}_{1.6}$ [m]\\ \colrule

        $\textrm{APR}$ & 4.03 & 0.37 & 0.67 &
        10.96 & 131.69 \\
                
        $\textrm{DD2}$ & 4.98 & 0.58 & 0.96 &
        5.48 & 188.16 \\
                
        $\textrm{DD2}_{3.3M_{\odot}}$ & 4.74 & 0.57 & 0.69 &
        13.63 & - \\
                
        $\textrm{NL3}$ & 6.01 & 0.64 & 0.97 &
        9.06 & 150.24 \\
                
        $\textrm{NL3}_{3.8M_{\odot}}$ & 6.13 & 0.64 & 0.95 &
        15.03 & - \\
                
        $\textrm{SFHo}$ & 3.46 & 0.25 & 0.75 &
        6.87 & 89.49 \\
                
        $\textrm{SFHo}_{3.2M_{\odot}}$ & 5.09 & 0.59 & 0.95 & - & -  \\
                
        $\textrm{SFHx}$ & 3.99 & 0.36 & 0.95 &
        3.83 & 242.07 \\
                
        $\textrm{Shen}$ & 5.43 & 0.62 & 0.96 &
        15.72 & 234.13 \\
                
        $\textrm{TM1}$ & 4.50 & 0.55 & 0.95 &
        10.05 & 175.38 \\
                
        $\textrm{TMa}$ & 4.75 & 0.57 & 0.97 &
        12.43 & 30.55 \\
                
        $\textrm{hlAPR}^{\dagger}$ & 10.53 & 2.15 & 0.97 &
        1.91 & 168.65 \\
                
        $\textrm{hsAPR}^{\dagger}$ & 4.13 & 0.43 & 0.77 &
        4.15 & 162.71 \\
                
        $\textrm{hsAPR}^{*}$ & 4.10 & 0.41 & 0.75 &
        9.55 & 59.79 \\
                
        $\textrm{hlAPR}^{*}$ & 8.05 & 1.11 & 0.96 &
        9.66 & 106.93 \\
                
        $\textrm{hlDD2}^{\dagger}$ & 12.17 & 3.21 & 0.99 &
        1.66 & 167.40 \\
                
        $\textrm{hsDD2}^{\dagger}$ & 6.68 & 0.71 & 0.98 &
        1.02 & 169.32 \\
                
        $\textrm{hlDD2}^{*}$ & 8.55 & 1.23 & 0.98 &
        12.40 & 242.25 \\
                
        $\textrm{hsDD2}^{*}$ & 5.36 & 0.61 & 0.96 &
        4.73 & 229.28 \\
                
\end{tabular}

\caption{Results summary showing effective range to which the CWB analysis is
sensitive and the expected detection rate assuming the ``realistic'' BNS
coalescence rate given in~\cite{ratesPaper}.
\emph{Classification accuracy} gives the probability that the post-merger
scenario (delayed vs prompt collapse) is correctly identified.
The delayed collapse waveforms are also characterised in terms of the median
error in the peak frequency measurement and, where appropriate, the median error
in the estimation of $R_{1.6}$\label{table:results}.}
\end{ruledtabular}
\end{table*}

\subsubsection{Waveform Classification \& Parameter Recovery}
\label{sec:waveform_classification}
\begin{figure}[ht]
\scalebox{1}{\includegraphics{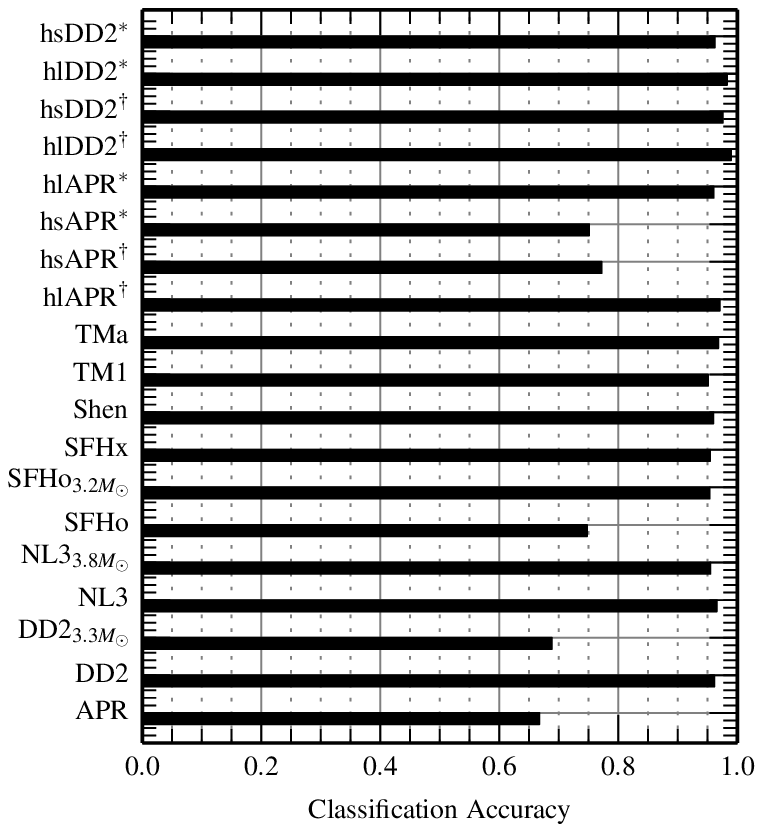}}
\caption{Classification accuracy: the probability that the post-merger scenario
is correctly identified.  Of these models, SFHo$_{3.2\msun}$ exhibits prompt
collapse to a black hole; all other simulations result in PMNS formation.
\label{fig:classification}}
\end{figure}

\begin{figure}
\centering
\subfloat[Part1][Median recovered frequencies]
{\scalebox{1}{\includegraphics{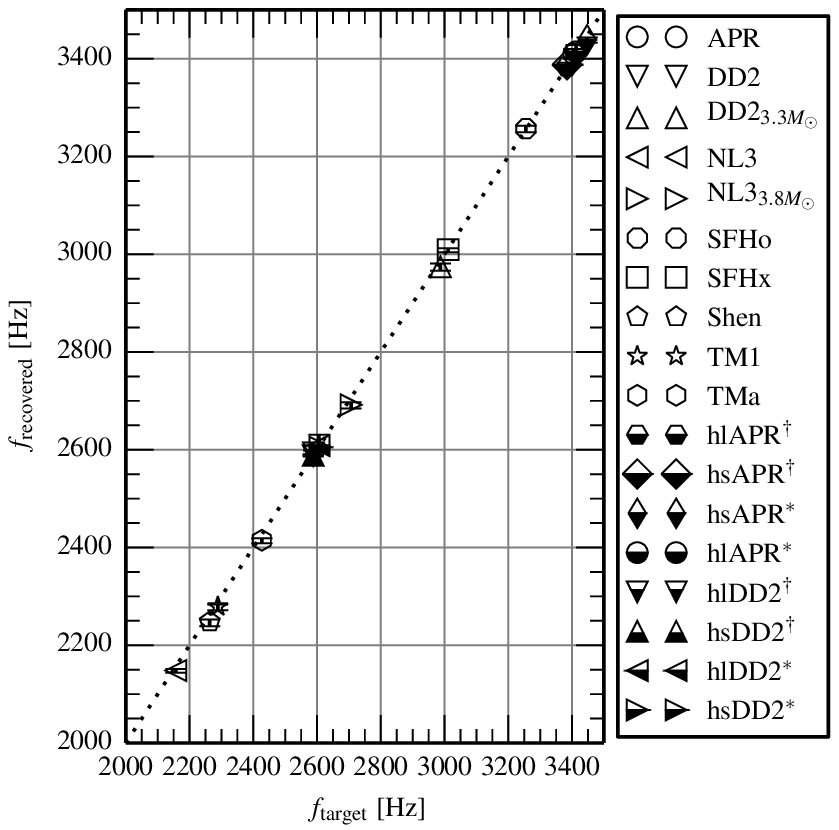}}\label{fig:frequency_estimates}}\\
\subfloat[Part2][Error in recovered frequencies]
{\scalebox{1}{\includegraphics{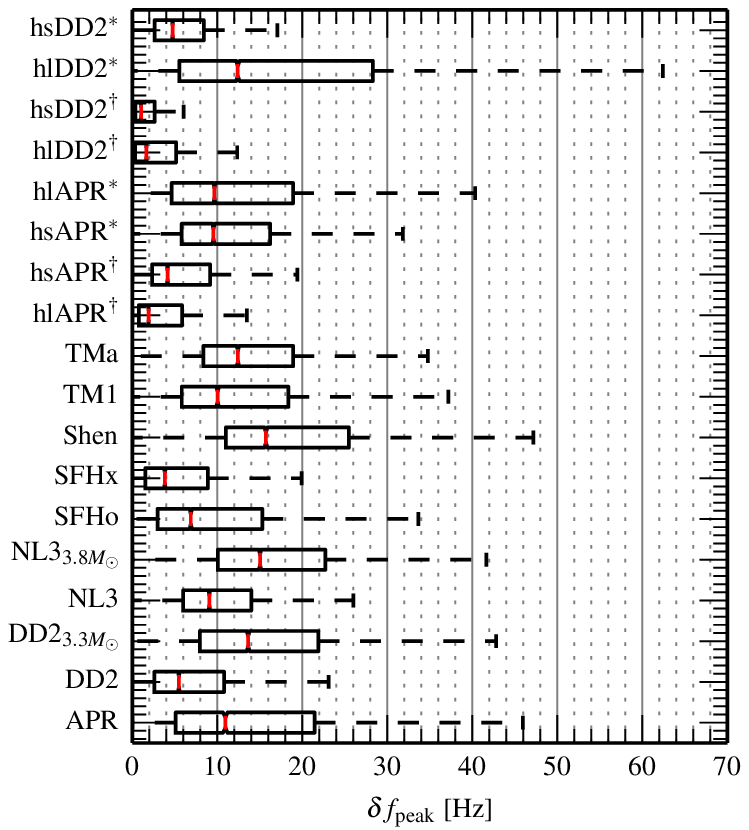}}\label{fig:frequency_errors}}
\caption{Frequency recovery for the PMNS waveforms.
\emph{Top panel}: the median recovered frequency as versus the target peak
frequency of the waveform.  Half-filled symbols indicate hybridized waveforms
(see sec~\ref{sec:hybrid_waveforms}).  \emph{Bottom panel}: the median (red
lines), interquartile ranges (boxes) and the minimum and maximum values within
1.5$\times$ the interquartile range of the absolute error in the frequency
determination for each waveform. \label{fig:frequency_recovery}}
\end{figure}

Figures~\ref{fig:classification},~\ref{fig:frequency_recovery}
and~\ref{fig:radius_recovery} summarize the results of 
this analysis,
marginalized over all extrinsic parameters, such as distance, sky location and
source orientation.
We characterize the performance of the classification scheme in terms of the
{\em classification accuracy}, which is the probability that the outcome of the
merger is correctly identified as prompt (SFHo$_{3.2\msun}$ simulation) or
delayed collapse (all other simulations).  This is evaluated as an efficiency
using equation~\ref{eq:efficiency}, where $k$ is now the number of correctly
classified signals and $N$ is the total number of detected signals.

In general, we find that the classification accuracy is better than 95\%; the
classification algorithm selects the correct post-merger scenario when
confronted with both PMNS and prompt-collapse waveforms.  We note, however,
that three of the PMNS waveforms studied, APR, DD2$_{3.2\msun}$ and SFHo, 
yield lower classification accuracies of $\sim 70\%$, 
as the SNR of the post-merger peak in these models comprises a
significantly lower fraction of the full waveform. In the cases
where the waveform is mis-classified, only the low frequency component of the
spectrum is loud enough to be reconstructed and no post-merger peak is visible. 

Figure~\ref{fig:frequency_recovery} summarizes the accuracy of the post-merger
frequency determination for our suite of PMNS waveforms.
Figure~\ref{fig:frequency_estimates} shows the recovered frequency
as a function of the target frequency for each waveform.
Figure~\ref{fig:frequency_errors} illustrates the accuracy of the frequency
measurement in terms of the median value and interquartile range of the
absolute deviation from the nominal target value:
\begin{equation}\label{eq:frequency_error}
\delta f_{\mathrm{peak}} \equiv \left| f_{\mathrm{peak}} -
f_{\mathrm{peak}}^{'}\right|.
\end{equation}
The median error lies in $\delta f_{\mathrm{peak}}\sim[4,15]$\,Hz for the
purely numerical waveforms and $\delta f_{\mathrm{peak}}\sim[2,12]$\,Hz for
the hybridized waveforms.  We find that the frequency measurements are most
accurate for waveforms with the most clearly defined and symmetric post-merger
peaks. 

\begin{figure}
\centering
\includegraphics{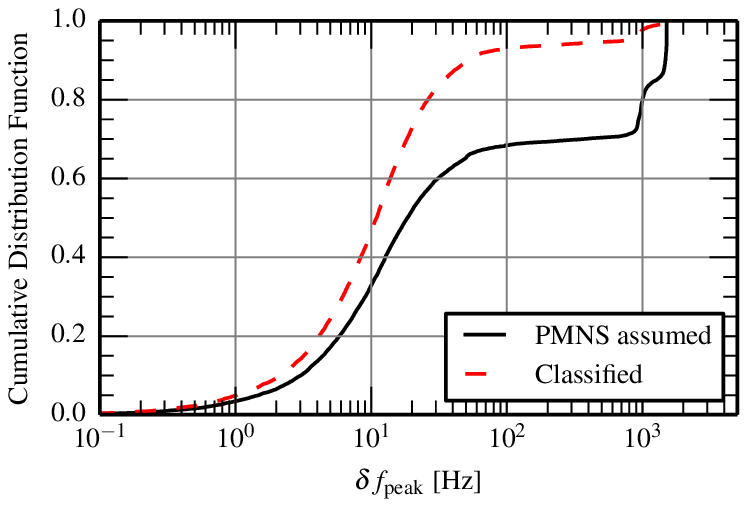}
\caption{The cumulative probability distribution of the absolute error in the
determination of peak frequency for APR waveforms.  The red, dashed trace shows
the frequency error after we have applied our classification scheme (see
section~\ref{sec:classification}).}
\label{fig:APR_deltaF_CDF}.
\end{figure}
\begin{figure}
\centering
\subfloat[Part1][Median recovered radii]
{\scalebox{1}{\includegraphics{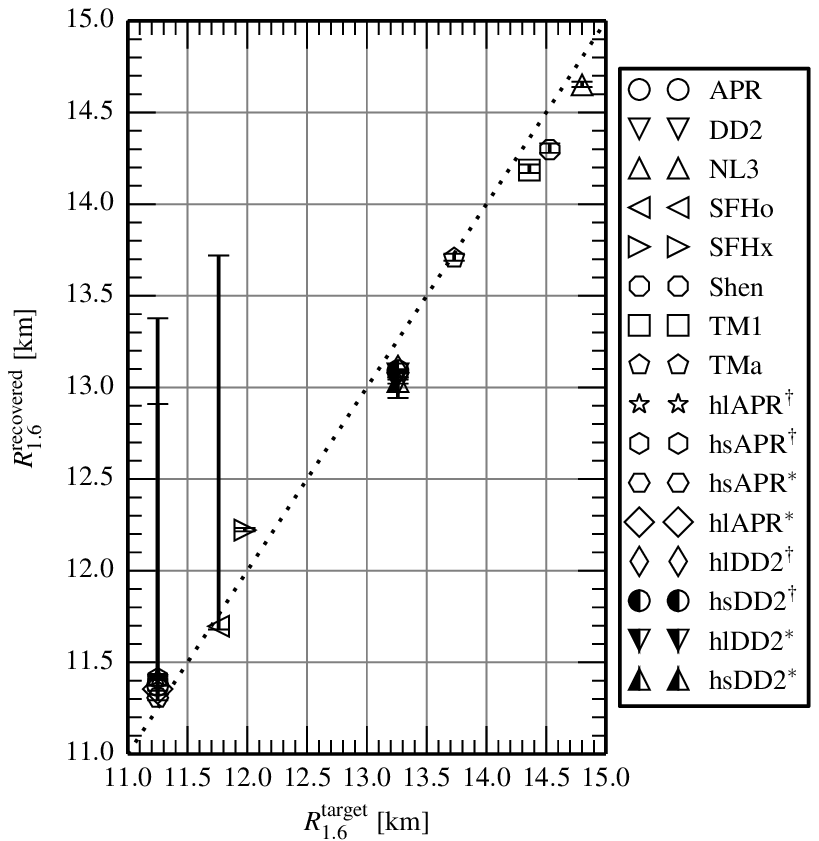}}\label{fig:radius_estimates}}\\
\subfloat[Part2][Error in recovered radii]
{\scalebox{1}{\includegraphics{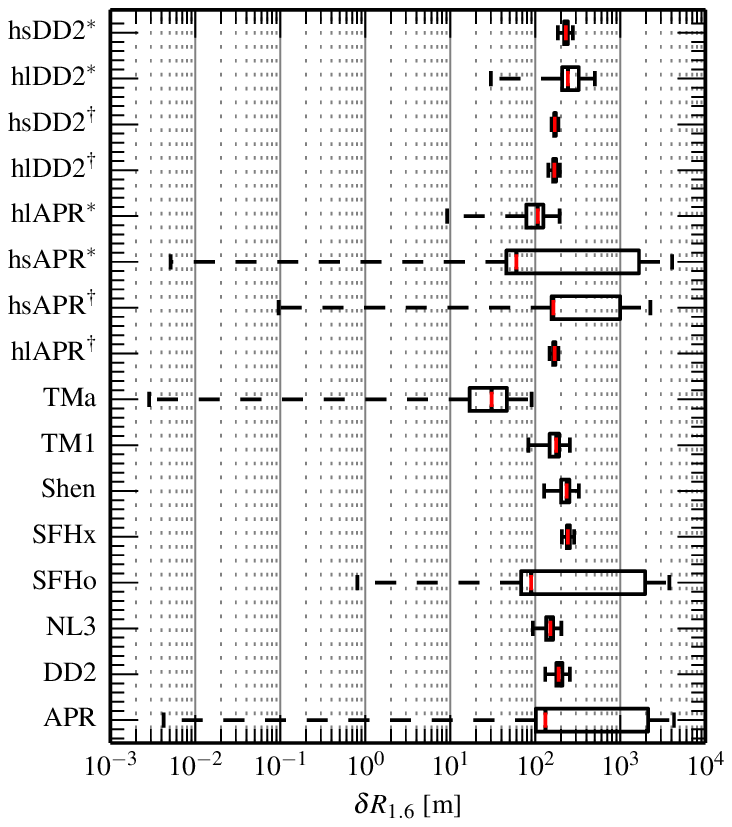}}\label{fig:radius_errors}}
\caption{Radius recovery for the waveforms exhibiting delayed collapse:
\emph{Top panel}: the median recovered radius versus the target radius for a
1.6$\msun$ star with that EoS.  \emph{Bottom panel}: the median (red lines),
interquartile ranges (boxes) and the minimum and maximum values within
1.5$\times$ the interquartile range of the absolute error in the radius
determination for each waveform. \label{fig:radius_recovery}}
\end{figure}

Given the relative likelihood that BNS coalescence results in the formation of a
PMNS for a wide variety of EoS and canonical NS masses
($m_1=m_2=1.35$\,M$_{\odot}$), it is natural to ask whether the classification
stage is necessary.  We find that the classification stage significantly
improves the robustness of the frequency estimation, particularly for waveforms
where the post-merger peak represents only a small fraction of the total power
in the waveform (i.e., APR, DD2$_{3.2\msun}$ and SFHo). 
To illustrate this, figure~\ref{fig:APR_deltaF_CDF} shows the cumulative
distribution of the frequency error $\delta f_{\mathrm{peak}}$ for the APR
waveform, with and without the classification procedure.  If we simply assume
that the merger results in PMNS formation, the best-fitting Gaussian component
for the spectrum (see equation~\ref{eq:gauss_plus_powlaw}) frequently lies at
much lower frequencies than the true post-merger peak, which may not be detected
and reconstructed by CWB at all, leading to serious errors in the the frequency
estimation.  Indeed, we find that the $90^{\mathrm{th}}$-percentile of the
frequency error with no classification stage is 1477\,kHz.  This value falls to
just 50\,Hz when we include the classification step.  We find similar results
for the DD2$_{3.2\msun}$ and SFHo waveforms.   The classification algorithm is,
therefore, an integral part of this analysis and helps to ensure that there is
reasonable evidence for the existence of the post-merger spectral peak prior to
estimating its frequency.

We conclude by considering the accuracy of the determination of the radius of a
reference $1.6\,\msun$ neutron star, using the measured $f_{\mathrm{peak}}^{'}$
and the fit of equation~\ref{eq:fpeak_radius} from~\cite{2012PhRvD..86f3001B}.
This fit is derived for systems with total mass $2.7\,\msun$. We thus restrict
this aspect of the analysis to those simulations with
M$_{\mathrm{tot}}=2.7\msun$.  It is worth noting that one can still expect a
correlation between $f_{\mathrm{peak}}$ and the radius of a reference neutron
star for different mass configurations but further systematic studies similar to
those in~\cite{2012PhRvD..86f3001B} will be necessary to obtain a fit for the
precise form of this relationship.  Figure~\ref{fig:radius_errors} shows the
distributions of the error in the measured radius, defined as:
\begin{equation}
\delta R_{1.6} = \left|R_{1.6} - R_{1.6}^{'} \right|,
\end{equation}
where $R_{1.6}^{'}$ is the radius from equation~\ref{eq:fpeak_radius} and the
measured post-merger frequency $f_{\mathrm{peak}}^{'}$.   We find the median
radius error lies in $\delta R_{1.6} \sim [30,250]$\,m, where the smallest
(largest) error is associated with the TMa (SFHx) waveform.

\section{Conclusion}
\label{sec:conclusion}
This paper presents the first systematic study of the expected detectability of
high-frequency bursts of \gw{s} from the merger and post-merger phase of binary
neutron star coalescence in the second generation of ground based detectors,
using the Coherent WaveBurst algorithm for unmodeled transient searches followed
by a classification scheme.

We determine the distance reach, and hence,
the expected detection rates for the CWB analysis through a large scale
Monte-Carlo study where simulated post-merger \gw{} signals are injected into
realistic non-Gaussian, non-stationary detector data which has been recolored
such that the noise power-spectral density matches the design goals of aLIGO and
AdV.

The results from the CWB Monte-Carlo study are compared with the expected range
and rates for a completely optimal matched-filter analysis assuming stationary,
Gaussian noise.   We find the effective range of the CWB analysis is
$\mathcal{R}_{\mathrm{CWB}} \sim 4\mbox{-}11$\,Mpc, depending on the
energy-content of the post-merger \gw{} signal, corresponding to an expected
detection rate $\dot{N}_{\mathrm{det}}^{\mathrm{CWB}} \sim
10^{-3}\mbox{-}0.1$\,events per year, assuming the `realistic' coalescence rate
$\dot{\mathcal{N}}=100$\,MWEG$^{-1}$\,Myr$^{-1}$ from~\cite{ratesPaper}.  If the
waveform is known exactly, permitting the use of an optimally-matched filter,
the post-merger signals considered in this paper may be detectable within a
sphere of radius $\mathcal{R}_{\mathrm{Opt}} \sim 13\mbox{-}27$\,Mpc, depending
on the energy-content of the post-merger \gw{} signal, corresponding to an
expected detection rate $\dot{N}_{\mathrm{det}}^{\mathrm{Opt}} \sim
0.03\mbox{-}0.3$\,events per year.  Both results assume that the threshold
required for detection corresponds to a false alarm probability of $p=10^{-5}$,
as required for a statistical significance of approximately $3\sigma$ for the
follow-up of $\mathcal{O}(100)$ events where the inspiral part of the \gw{}
signal has been detected.   

While there is nearly an order of magnitude difference between the sensitivity
of the CWB analysis and the expectation for the idealised matched-filter, it is
important to stress that the optimal sensitivity is highly unlikely to be
realised in practice;  even if there was a sufficiently accurate analytic form
for the merger/post-merger signal which would facilitate the construction of a
matched-filter, the start time of the signal, the sky-location and the intrinsic
parameters of the source would still be unknown.  Searching over the unknown
parameter space introduces a trials factor into the SNR distribution, reducing
the sensitivity of the search.  Furthermore, real detector data is rarely
Gaussian or stationary, which tends to increase the threshold required for
detection futher.  The CWB analysis, by contrast, makes no assumptions regarding
waveform morphology, uses an existing and well-tested data analysis pipeline and
uses recoloured, initial-detector data which may plausibly be regarded to share
the characteristics of advanced detector data.  The optimal and CWB estimates
may therefore be regarded as absolute upper and realistic lower bounds on the
detectability of high-frequency \gw{} signals from binary neutron star
coalescence keeping in mind that the strength of the \gw{} emission is not
exactly known from numerical simulations.

We have also developed and demonstrated the efficacy of a simple model-selection
and parameter estimation algorithm which distinguishes between the post-merger
scenarios of prompt- and delayed-collapse to a black hole.  This procedure,
which has negligible computational cost as compared with the CWB analysis
itself, assumes that any statistically significant high-frequency signal power
following a binary neutron star inspiral is due to the merger/post-merger \gw{}
emission from the coalescence.  For delayed collapse, we expect the
reconstructed waveform to resemble a power-law decay with a Gaussian peak
somewhere in $\sim [2,~4]$\,kHz, while no such peak is expected in the case of
prompt-collapse to a black hole.  We deploy the Bayesian Information Criterion
to select between these models and find that, for most of the waveforms
considered in this study, the probability of correctly identifying the
post-merger scenario is greater than $95\%$.  Delayed collapse waveforms in
which the post-merger spectral peak comprises a smaller fraction of the total
SNR prove harder to correctly classify since the spectrum is dominated by power
from the late inspiral and merger as is the case with prompt collapse.

When the outcome of the merger is identified as delayed-collapse, the
reconstruction analysis also returns the maximum-likelihood estimate of the
center frequency for the Gaussian peak which we identify with the dominant
post-merger oscillation frequency.  We find that the typical magnitude of the
error in this determination of the peak frequency is $\tilde{\delta
f}_{\mathrm{peak}} \sim 1\mbox{-}10$\,Hz, with the highest accuracy
corresponding to waveforms in which the SNR is dominated by the contribution
from the post-merger oscillations.   In addition, the model selection stage used
to distinguish prompt and delayed collapse increases the robustness of this
frequency estimation by ensuring that there is indeed evidence for post-merger
oscillations before attempting to measure their dominant frequency content. 

Finally, following~\cite{2012PhRvD..86f3001B}, we use the measured peak
frequency to estimate the radius of a $1.6~\msun$ neutron star and compare the
result to the true radius of such a star with each waveform's equation of state.
Using the fit described in~\cite{2012PhRvD..86f3001B} (i.e.,
equation~\ref{eq:fpeak_radius}), we find that the typical magnitude of the error
in this radius estimate is $\tilde{\delta R}_{1.6} \sim 100\mbox{-}200$\,m,
where the dominant source of error is in the fit itself associated with the
scatter of the fitted model results rather than the measurement uncertainty.  
As remarked upon in section~\ref{sec:waveform_classification}, these estimates
for the error in the recovered radius are based on the fit for a specific set of
simulations (symmetric mass configuration with a total mass of $2.7\msun$) where
the $f_{\mathrm{peak}}\mbox{-}R_{1.6}$ relationship has  been carefully studied.
Further surveys of the post-merger waveform, similar to that
in~\cite{2012PhRvD..86f3001B}, will be important to obtain relationships similar
to the $f_{\mathrm{peak}}\mbox{-}R_{1.6}$ correlation for a variety of mass
configurations and to account for different approaches and detailed physics used
in various modelling codes.


We see then that the prospects for high-frequency searches for the post-merger
signal following BNS coalescence rely on serendipitous nearby events and
optimistic coalescence rates.  While such a scenario may be unlikely, even in
the advanced detector era, it is difficult to overstate the rewards of the
detection and characterisation of the post-merger signal.  Indeed, as we have
shown in this work, the combination of very simple modelling of the signal
spectrum and existing data analysis techniques allows one to correctly identify
the post-merger scenario and, in the case of delayed collapse, accurately
measure the dominant post-merger oscillation frequency.  Furthermore,
refinements and advances in data analysis ranging from modifications such as
improved time-frequency clustering and choice of basis for the CWB algorithm
used in this study, to novel Bayesian techniques for the robust detection and
characterisation of un-modelled signals
(e.g.,~\cite{2010PhRvD..82j3007L,2012arXiv1204.2000C}) may lead to
significant improvements in the prospects for detecting the high-frequency \gw{}
emission following binary neutron star coalescence.  In particular, we consider
the development of analytic templates for the post-merger signal such as
in~\cite{2013PhRvD..88d4026H} and their deployment using Bayesian inference
algorithms such as those described in~\cite{2013PhRvD..88f2001A} of paramount
importance and a high-priority goal for the follow-up of the immiment first
detections of binary neutron star inspiral signals in the advanced detector era.

\acknowledgments

The authors thank Francesco Pannarale for useful input and careful reading of
this manuscript and S. Klimenko, I.S. Heng and M. West for helpful
exchanges and discussion.
J.C. and L.C. gratefully acknowledge support from NSF grant PHY-0955773.
C.P. gratefully acknowledges NSF grants PHY-0970074, PHY-1307429 and the UWM
Research Growth Initiative.
A.B. is a Marie Curie Intra-European Fellow within the 7th European Community
Framework Programme (IEF 331873). This work was supported by the Deutsche
Forschungsgemeinschaft through Sonderforschungsbe\-reich Transregio 7
``Gravitational Wave Astronomy'', and the Cluster of Excellence EXC 153 ``Origin
and Structure of the Universe''.  Partial support comes from ``NewCompStar'',
COST Action MP1304. The computations were performed at the Rechenzentrum
Garching of the Max-Planck-Gesellschaft, the Max Planck Institute for
Astrophysics, and the Cyprus Institute under the LinkSCEEM/Cy-Tera project.

\bibliographystyle{apsrev}
\bibliography{postmerger_gws}

\begin{thebibliography}{108}
\expandafter\ifx\csname natexlab\endcsname\relax\def\natexlab#1{#1}\fi
\expandafter\ifx\csname bibnamefont\endcsname\relax
  \def\bibnamefont#1{#1}\fi
\expandafter\ifx\csname bibfnamefont\endcsname\relax
  \def\bibfnamefont#1{#1}\fi
\expandafter\ifx\csname citenamefont\endcsname\relax
  \def\citenamefont#1{#1}\fi
\expandafter\ifx\csname url\endcsname\relax
  \def\url#1{\texttt{#1}}\fi
\expandafter\ifx\csname urlprefix\endcsname\relax\def\urlprefix{URL }\fi
\providecommand{\bibinfo}[2]{#2}
\providecommand{\eprint}[2][]{\url{#2}}

\bibitem[{\citenamefont{{Harry} et~al.}(2010)}]{2010CQGra..27h4006H}
\bibinfo{author}{\bibfnamefont{G.~M.} \bibnamefont{{Harry}}}
  \bibnamefont{et~al.} (\bibinfo{collaboration}{LIGO Scientific
  Collaboration}), \bibinfo{journal}{Class. Quant. Grav.}
  \textbf{\bibinfo{volume}{27}}, \bibinfo{eid}{084006} (\bibinfo{year}{2010}).

\bibitem[{\citenamefont{{Accadia} et~al.}(2011)}]{2011CQGra..28k4002A}
\bibinfo{author}{\bibfnamefont{T.}~\bibnamefont{{Accadia}}}
  \bibnamefont{et~al.}, \bibinfo{journal}{Class. Quant. Grav.}
  \textbf{\bibinfo{volume}{28}}, \bibinfo{eid}{114002} (\bibinfo{year}{2011}).

\bibitem[{\citenamefont{Collaboration}(2009)}]{virgo_baseline}
\bibinfo{author}{\bibfnamefont{T.~V.} \bibnamefont{Collaboration}},
  \bibinfo{journal}{Virgo Tech. Rep. VIR-027A-09}  (\bibinfo{year}{2009}),
  \urlprefix\url{https://tds.ego-gw.it/itf/tds/file.php?callFile=VIR-0027A-09.pdf}.

\bibitem[{\citenamefont{{Kuroda}}(2010)}]{2010CQGra..27h4004K}
\bibinfo{author}{\bibfnamefont{K.}~\bibnamefont{{Kuroda}}}
  (\bibinfo{collaboration}{LCGT Collaboration}), \bibinfo{journal}{Class.
  Quant. Grav.} \textbf{\bibinfo{volume}{27}}, \bibinfo{eid}{084004}
  (\bibinfo{year}{2010}).

\bibitem[{\citenamefont{{Aasi}
  et~al.}(2013{\natexlab{a}})}]{2013arXiv1304.0670L}
\bibinfo{author}{\bibfnamefont{J.}~\bibnamefont{{Aasi}}} \bibnamefont{et~al.}
  (\bibinfo{collaboration}{{LIGO Scientific Collaboration} and {Virgo
  Collaboration}}), \bibinfo{journal}{ArXiv e-prints}
  (\bibinfo{year}{2013}{\natexlab{a}}), \eprint{1304.0670}.

\bibitem[{\citenamefont{{Abadie} et~al.}(2010)}]{ratesPaper}
\bibinfo{author}{\bibfnamefont{J.}~\bibnamefont{{Abadie}}} \bibnamefont{et~al.}
  (\bibinfo{collaboration}{{LIGO Scientific Collaboration} and {Virgo
  Collaboration}}), \bibinfo{journal}{Classical and Quantum Gravity}
  \textbf{\bibinfo{volume}{27}}, \bibinfo{pages}{173001}
  (\bibinfo{year}{2010}),
  \urlprefix\url{http://stacks.iop.org/0264-9381/27/i=17/a=173001}.

\bibitem[{\citenamefont{{Lattimer} and {Prakash}}(2007)}]{2007PhR...442..109L}
\bibinfo{author}{\bibfnamefont{J.~M.} \bibnamefont{{Lattimer}}}
  \bibnamefont{and}
  \bibinfo{author}{\bibfnamefont{M.}~\bibnamefont{{Prakash}}},
  \bibinfo{journal}{Physics Reports} \textbf{\bibinfo{volume}{442}},
  \bibinfo{pages}{109} (\bibinfo{year}{2007}), \eprint{astro-ph/0612440}.

\bibitem[{\citenamefont{{Flanagan} and {Hinderer}}(2008)}]{2008PhRvD..77b1502F}
\bibinfo{author}{\bibfnamefont{{\'E}.~{\'E}.} \bibnamefont{{Flanagan}}}
  \bibnamefont{and}
  \bibinfo{author}{\bibfnamefont{T.}~\bibnamefont{{Hinderer}}},
  \bibinfo{journal}{\prd} \textbf{\bibinfo{volume}{77}}, \bibinfo{eid}{021502}
  (\bibinfo{year}{2008}), \eprint{0709.1915}.

\bibitem[{\citenamefont{{Baiotti} et~al.}(2010)\citenamefont{{Baiotti},
  {Damour}, {Giacomazzo}, {Nagar}, and {Rezzolla}}}]{2010PhRvL.105z1101B}
\bibinfo{author}{\bibfnamefont{L.}~\bibnamefont{{Baiotti}}},
  \bibinfo{author}{\bibfnamefont{T.}~\bibnamefont{{Damour}}},
  \bibinfo{author}{\bibfnamefont{B.}~\bibnamefont{{Giacomazzo}}},
  \bibinfo{author}{\bibfnamefont{A.}~\bibnamefont{{Nagar}}}, \bibnamefont{and}
  \bibinfo{author}{\bibfnamefont{L.}~\bibnamefont{{Rezzolla}}},
  \bibinfo{journal}{Phys. Rev. Lett.} \textbf{\bibinfo{volume}{105}},
  \bibinfo{eid}{261101} (\bibinfo{year}{2010}), \eprint{1009.0521}.

\bibitem[{\citenamefont{{Bernuzzi} et~al.}(2012)\citenamefont{{Bernuzzi},
  {Nagar}, {Thierfelder}, and {Br{\"u}gmann}}}]{2012PhRvD..86d4030B}
\bibinfo{author}{\bibfnamefont{S.}~\bibnamefont{{Bernuzzi}}},
  \bibinfo{author}{\bibfnamefont{A.}~\bibnamefont{{Nagar}}},
  \bibinfo{author}{\bibfnamefont{M.}~\bibnamefont{{Thierfelder}}},
  \bibnamefont{and}
  \bibinfo{author}{\bibfnamefont{B.}~\bibnamefont{{Br{\"u}gmann}}},
  \bibinfo{journal}{\prd} \textbf{\bibinfo{volume}{86}}, \bibinfo{eid}{044030}
  (\bibinfo{year}{2012}), \eprint{1205.3403}.

\bibitem[{\citenamefont{{Read} et~al.}(2009)\citenamefont{{Read}, {Markakis},
  {Shibata}, {Ury{\= u}}, {Creighton}, and {Friedman}}}]{2009PhRvD..79l4033R}
\bibinfo{author}{\bibfnamefont{J.~S.} \bibnamefont{{Read}}},
  \bibinfo{author}{\bibfnamefont{C.}~\bibnamefont{{Markakis}}},
  \bibinfo{author}{\bibfnamefont{M.}~\bibnamefont{{Shibata}}},
  \bibinfo{author}{\bibfnamefont{K.}~\bibnamefont{{Ury{\= u}}}},
  \bibinfo{author}{\bibfnamefont{J.~D.~E.} \bibnamefont{{Creighton}}},
  \bibnamefont{and} \bibinfo{author}{\bibfnamefont{J.~L.}
  \bibnamefont{{Friedman}}}, \bibinfo{journal}{\prd}
  \textbf{\bibinfo{volume}{79}}, \bibinfo{eid}{124033} (\bibinfo{year}{2009}),
  \eprint{0901.3258}.

\bibitem[{\citenamefont{{Zhuge} et~al.}(1994)\citenamefont{{Zhuge},
  {Centrella}, and {McMillan}}}]{1994PhRvD..50.6247Z}
\bibinfo{author}{\bibfnamefont{X.}~\bibnamefont{{Zhuge}}},
  \bibinfo{author}{\bibfnamefont{J.~M.} \bibnamefont{{Centrella}}},
  \bibnamefont{and} \bibinfo{author}{\bibfnamefont{S.~L.~W.}
  \bibnamefont{{McMillan}}}, \bibinfo{journal}{\prd}
  \textbf{\bibinfo{volume}{50}}, \bibinfo{pages}{6247} (\bibinfo{year}{1994}),
  \eprint{gr-qc/9411029}.

\bibitem[{\citenamefont{{Ruffert} et~al.}(1996)\citenamefont{{Ruffert},
  {Janka}, and {Schaefer}}}]{1996A&A...311..532R}
\bibinfo{author}{\bibfnamefont{M.}~\bibnamefont{{Ruffert}}},
  \bibinfo{author}{\bibfnamefont{H.-T.} \bibnamefont{{Janka}}},
  \bibnamefont{and}
  \bibinfo{author}{\bibfnamefont{G.}~\bibnamefont{{Schaefer}}},
  \bibinfo{journal}{Astron. Astrophys.} \textbf{\bibinfo{volume}{311}},
  \bibinfo{pages}{532} (\bibinfo{year}{1996}), \eprint{astro-ph/9509006}.

\bibitem[{\citenamefont{{Shibata}}(2005)}]{2005PhRvL..94t1101S}
\bibinfo{author}{\bibfnamefont{M.}~\bibnamefont{{Shibata}}},
  \bibinfo{journal}{Phys. Rev. Lett.} \textbf{\bibinfo{volume}{94}},
  \bibinfo{eid}{201101} (\bibinfo{year}{2005}), \eprint{gr-qc/0504082}.

\bibitem[{\citenamefont{{Shibata} et~al.}(2005)\citenamefont{{Shibata},
  {Taniguchi}, and {Ury{\= u}}}}]{2005PhRvD..71h4021S}
\bibinfo{author}{\bibfnamefont{M.}~\bibnamefont{{Shibata}}},
  \bibinfo{author}{\bibfnamefont{K.}~\bibnamefont{{Taniguchi}}},
  \bibnamefont{and} \bibinfo{author}{\bibfnamefont{K.}~\bibnamefont{{Ury{\=
  u}}}}, \bibinfo{journal}{\prd} \textbf{\bibinfo{volume}{71}},
  \bibinfo{eid}{084021} (\bibinfo{year}{2005}), \eprint{gr-qc/0503119}.

\bibitem[{\citenamefont{{Oechslin} et~al.}(2007)\citenamefont{{Oechslin},
  {Janka}, and {Marek}}}]{2007A&A...467..395O}
\bibinfo{author}{\bibfnamefont{R.}~\bibnamefont{{Oechslin}}},
  \bibinfo{author}{\bibfnamefont{H.-T.} \bibnamefont{{Janka}}},
  \bibnamefont{and} \bibinfo{author}{\bibfnamefont{A.}~\bibnamefont{{Marek}}},
  \bibinfo{journal}{Astron. Astrophys.} \textbf{\bibinfo{volume}{467}},
  \bibinfo{pages}{395} (\bibinfo{year}{2007}).

\bibitem[{\citenamefont{{Oechslin} and {Janka}}(2007)}]{2007PhRvL..99l1102O}
\bibinfo{author}{\bibfnamefont{R.}~\bibnamefont{{Oechslin}}} \bibnamefont{and}
  \bibinfo{author}{\bibfnamefont{H.-T.} \bibnamefont{{Janka}}},
  \bibinfo{journal}{Phys. Rev. Lett.} \textbf{\bibinfo{volume}{99}},
  \bibinfo{eid}{121102} (\bibinfo{year}{2007}), \eprint{astro-ph/0702228}.

\bibitem[{\citenamefont{{Anderson} et~al.}(2008)\citenamefont{{Anderson},
  {Hirschmann}, {Lehner}, {Liebling}, {Motl}, {Neilsen}, {Palenzuela}, and
  {Tohline}}}]{2008PhRvD..77b4006A}
\bibinfo{author}{\bibfnamefont{M.}~\bibnamefont{{Anderson}}},
  \bibinfo{author}{\bibfnamefont{E.~W.} \bibnamefont{{Hirschmann}}},
  \bibinfo{author}{\bibfnamefont{L.}~\bibnamefont{{Lehner}}},
  \bibinfo{author}{\bibfnamefont{S.~L.} \bibnamefont{{Liebling}}},
  \bibinfo{author}{\bibfnamefont{P.~M.} \bibnamefont{{Motl}}},
  \bibinfo{author}{\bibfnamefont{D.}~\bibnamefont{{Neilsen}}},
  \bibinfo{author}{\bibfnamefont{C.}~\bibnamefont{{Palenzuela}}},
  \bibnamefont{and} \bibinfo{author}{\bibfnamefont{J.~E.}
  \bibnamefont{{Tohline}}}, \bibinfo{journal}{\prd}
  \textbf{\bibinfo{volume}{77}}, \bibinfo{eid}{024006} (\bibinfo{year}{2008}),
  \eprint{0708.2720}.

\bibitem[{\citenamefont{{Liu} et~al.}(2008)\citenamefont{{Liu}, {Shapiro},
  {Etienne}, and {Taniguchi}}}]{2008PhRvD..78b4012L}
\bibinfo{author}{\bibfnamefont{Y.~T.} \bibnamefont{{Liu}}},
  \bibinfo{author}{\bibfnamefont{S.~L.} \bibnamefont{{Shapiro}}},
  \bibinfo{author}{\bibfnamefont{Z.~B.} \bibnamefont{{Etienne}}},
  \bibnamefont{and}
  \bibinfo{author}{\bibfnamefont{K.}~\bibnamefont{{Taniguchi}}},
  \bibinfo{journal}{\prd} \textbf{\bibinfo{volume}{78}}, \bibinfo{eid}{024012}
  (\bibinfo{year}{2008}), \eprint{0803.4193}.

\bibitem[{\citenamefont{{Baiotti} et~al.}(2008)\citenamefont{{Baiotti},
  {Giacomazzo}, and {Rezzolla}}}]{PhysRevD.78.084033}
\bibinfo{author}{\bibfnamefont{L.}~\bibnamefont{{Baiotti}}},
  \bibinfo{author}{\bibfnamefont{B.}~\bibnamefont{{Giacomazzo}}},
  \bibnamefont{and}
  \bibinfo{author}{\bibfnamefont{L.}~\bibnamefont{{Rezzolla}}},
  \bibinfo{journal}{Phys. Rev. D} \textbf{\bibinfo{volume}{78}},
  \bibinfo{pages}{084033} (\bibinfo{year}{2008}),
  \urlprefix\url{http://link.aps.org/doi/10.1103/PhysRevD.78.084033}.

\bibitem[{\citenamefont{{Kiuchi} et~al.}(2009)\citenamefont{{Kiuchi},
  {Sekiguchi}, {Shibata}, and {Taniguchi}}}]{2009PhRvD..80f4037K}
\bibinfo{author}{\bibfnamefont{K.}~\bibnamefont{{Kiuchi}}},
  \bibinfo{author}{\bibfnamefont{Y.}~\bibnamefont{{Sekiguchi}}},
  \bibinfo{author}{\bibfnamefont{M.}~\bibnamefont{{Shibata}}},
  \bibnamefont{and}
  \bibinfo{author}{\bibfnamefont{K.}~\bibnamefont{{Taniguchi}}},
  \bibinfo{journal}{\prd} \textbf{\bibinfo{volume}{80}}, \bibinfo{eid}{064037}
  (\bibinfo{year}{2009}), \eprint{0904.4551}.

\bibitem[{\citenamefont{{Stergioulas} et~al.}(2011)\citenamefont{{Stergioulas},
  {Bauswein}, {Zagkouris}, and {Janka}}}]{2011MNRAS.418..427S}
\bibinfo{author}{\bibfnamefont{N.}~\bibnamefont{{Stergioulas}}},
  \bibinfo{author}{\bibfnamefont{A.}~\bibnamefont{{Bauswein}}},
  \bibinfo{author}{\bibfnamefont{K.}~\bibnamefont{{Zagkouris}}},
  \bibnamefont{and} \bibinfo{author}{\bibfnamefont{H.-T.}
  \bibnamefont{{Janka}}}, \bibinfo{journal}{Mon. Not. Roy. Astron. Soc.}
  \textbf{\bibinfo{volume}{418}}, \bibinfo{pages}{427} (\bibinfo{year}{2011}).

\bibitem[{\citenamefont{{Giacomazzo} et~al.}(2011)\citenamefont{{Giacomazzo},
  {Rezzolla}, and {Baiotti}}}]{2011PhRvD..83d4014G}
\bibinfo{author}{\bibfnamefont{B.}~\bibnamefont{{Giacomazzo}}},
  \bibinfo{author}{\bibfnamefont{L.}~\bibnamefont{{Rezzolla}}},
  \bibnamefont{and}
  \bibinfo{author}{\bibfnamefont{L.}~\bibnamefont{{Baiotti}}},
  \bibinfo{journal}{\prd} \textbf{\bibinfo{volume}{83}}, \bibinfo{eid}{044014}
  (\bibinfo{year}{2011}), \eprint{1009.2468}.

\bibitem[{\citenamefont{{Hotokezaka} et~al.}(2011)\citenamefont{{Hotokezaka},
  {Kyutoku}, {Okawa}, {Shibata}, and {Kiuchi}}}]{2011PhRvD..83l4008H}
\bibinfo{author}{\bibfnamefont{K.}~\bibnamefont{{Hotokezaka}}},
  \bibinfo{author}{\bibfnamefont{K.}~\bibnamefont{{Kyutoku}}},
  \bibinfo{author}{\bibfnamefont{H.}~\bibnamefont{{Okawa}}},
  \bibinfo{author}{\bibfnamefont{M.}~\bibnamefont{{Shibata}}},
  \bibnamefont{and} \bibinfo{author}{\bibfnamefont{K.}~\bibnamefont{{Kiuchi}}},
  \bibinfo{journal}{\prd} \textbf{\bibinfo{volume}{83}}, \bibinfo{eid}{124008}
  (\bibinfo{year}{2011}), \eprint{1105.4370}.

\bibitem[{\citenamefont{{Sekiguchi} et~al.}(2011)\citenamefont{{Sekiguchi},
  {Kiuchi}, {Kyutoku}, and {Shibata}}}]{2011PhRvL.107e1102S}
\bibinfo{author}{\bibfnamefont{Y.}~\bibnamefont{{Sekiguchi}}},
  \bibinfo{author}{\bibfnamefont{K.}~\bibnamefont{{Kiuchi}}},
  \bibinfo{author}{\bibfnamefont{K.}~\bibnamefont{{Kyutoku}}},
  \bibnamefont{and}
  \bibinfo{author}{\bibfnamefont{M.}~\bibnamefont{{Shibata}}},
  \bibinfo{journal}{prl} \textbf{\bibinfo{volume}{107}}, \bibinfo{eid}{051102}
  (\bibinfo{year}{2011}), \eprint{1105.2125}.

\bibitem[{\citenamefont{{Bauswein} and {Janka}}(2012)}]{2012PhRvL.108a1101B}
\bibinfo{author}{\bibfnamefont{A.}~\bibnamefont{{Bauswein}}} \bibnamefont{and}
  \bibinfo{author}{\bibfnamefont{H.-T.} \bibnamefont{{Janka}}},
  \bibinfo{journal}{Physical Review Letters} \textbf{\bibinfo{volume}{108}},
  \bibinfo{eid}{011101} (\bibinfo{year}{2012}), \eprint{1106.1616}.

\bibitem[{\citenamefont{{Bauswein}
  et~al.}(2012{\natexlab{a}})\citenamefont{{Bauswein}, {Janka}, {Hebeler}, and
  {Schwenk}}}]{2012PhRvD..86f3001B}
\bibinfo{author}{\bibfnamefont{A.}~\bibnamefont{{Bauswein}}},
  \bibinfo{author}{\bibfnamefont{H.-T.} \bibnamefont{{Janka}}},
  \bibinfo{author}{\bibfnamefont{K.}~\bibnamefont{{Hebeler}}},
  \bibnamefont{and}
  \bibinfo{author}{\bibfnamefont{A.}~\bibnamefont{{Schwenk}}},
  \bibinfo{journal}{\prd} \textbf{\bibinfo{volume}{86}}, \bibinfo{eid}{063001}
  (\bibinfo{year}{2012}{\natexlab{a}}).

\bibitem[{\citenamefont{{Bauswein} et~al.}(2013)\citenamefont{{Bauswein},
  {Baumgarte}, and {Janka}}}]{2013PhRvL.111m1101B}
\bibinfo{author}{\bibfnamefont{A.}~\bibnamefont{{Bauswein}}},
  \bibinfo{author}{\bibfnamefont{T.~W.} \bibnamefont{{Baumgarte}}},
  \bibnamefont{and} \bibinfo{author}{\bibfnamefont{H.-T.}
  \bibnamefont{{Janka}}}, \bibinfo{journal}{\prl}
  \textbf{\bibinfo{volume}{111}}, \bibinfo{eid}{131101} (\bibinfo{year}{2013}).

\bibitem[{\citenamefont{{Hotokezaka} et~al.}(2013)\citenamefont{{Hotokezaka},
  {Kiuchi}, {Kyutoku}, {Muranushi}, {Sekiguchi}, {Shibata}, and
  {Taniguchi}}}]{2013PhRvD..88d4026H}
\bibinfo{author}{\bibfnamefont{K.}~\bibnamefont{{Hotokezaka}}},
  \bibinfo{author}{\bibfnamefont{K.}~\bibnamefont{{Kiuchi}}},
  \bibinfo{author}{\bibfnamefont{K.}~\bibnamefont{{Kyutoku}}},
  \bibinfo{author}{\bibfnamefont{T.}~\bibnamefont{{Muranushi}}},
  \bibinfo{author}{\bibfnamefont{Y.}~\bibnamefont{{Sekiguchi}}},
  \bibinfo{author}{\bibfnamefont{M.}~\bibnamefont{{Shibata}}},
  \bibnamefont{and}
  \bibinfo{author}{\bibfnamefont{K.}~\bibnamefont{{Taniguchi}}},
  \bibinfo{journal}{\prd} \textbf{\bibinfo{volume}{88}}, \bibinfo{eid}{044026}
  (\bibinfo{year}{2013}).

\bibitem[{\citenamefont{{Bernuzzi} et~al.}(2013)\citenamefont{{Bernuzzi},
  {Dietrich}, {Tichy}, and {Bruegmann}}}]{2013arXiv1311.4443B}
\bibinfo{author}{\bibfnamefont{S.}~\bibnamefont{{Bernuzzi}}},
  \bibinfo{author}{\bibfnamefont{T.}~\bibnamefont{{Dietrich}}},
  \bibinfo{author}{\bibfnamefont{W.}~\bibnamefont{{Tichy}}}, \bibnamefont{and}
  \bibinfo{author}{\bibfnamefont{B.}~\bibnamefont{{Bruegmann}}},
  \bibinfo{journal}{ArXiv e-prints}  (\bibinfo{year}{2013}),
  \eprint{1311.4443}.

\bibitem[{\citenamefont{{Takami} et~al.}(2014)\citenamefont{{Takami},
  {Rezzolla}, and {Baiotti}}}]{2014arXiv1403.5672T}
\bibinfo{author}{\bibfnamefont{K.}~\bibnamefont{{Takami}}},
  \bibinfo{author}{\bibfnamefont{L.}~\bibnamefont{{Rezzolla}}},
  \bibnamefont{and}
  \bibinfo{author}{\bibfnamefont{L.}~\bibnamefont{{Baiotti}}},
  \bibinfo{journal}{ArXiv e-prints}  (\bibinfo{year}{2014}),
  \eprint{1403.5672}.

\bibitem[{\citenamefont{{Baumgarte} et~al.}(2000)\citenamefont{{Baumgarte},
  {Shapiro}, and {Shibata}}}]{2000ApJ...528L..29B}
\bibinfo{author}{\bibfnamefont{T.~W.} \bibnamefont{{Baumgarte}}},
  \bibinfo{author}{\bibfnamefont{S.~L.} \bibnamefont{{Shapiro}}},
  \bibnamefont{and}
  \bibinfo{author}{\bibfnamefont{M.}~\bibnamefont{{Shibata}}},
  \bibinfo{journal}{Astrophys. J. Lett.} \textbf{\bibinfo{volume}{528}},
  \bibinfo{pages}{L29} (\bibinfo{year}{2000}), \eprint{astro-ph/9910565}.

\bibitem[{\citenamefont{{Bauswein} et~al.}(2014)\citenamefont{{Bauswein},
  {Stergioulas}, and {Janka}}}]{2014arXiv1403.5301B}
\bibinfo{author}{\bibfnamefont{A.}~\bibnamefont{{Bauswein}}},
  \bibinfo{author}{\bibfnamefont{N.}~\bibnamefont{{Stergioulas}}},
  \bibnamefont{and} \bibinfo{author}{\bibfnamefont{H.-T.}
  \bibnamefont{{Janka}}}, \bibinfo{journal}{ArXiv e-prints}
  (\bibinfo{year}{2014}), \eprint{1403.5301}.

\bibitem[{\citenamefont{{Punturo} et~al.}(2010)\citenamefont{{Punturo},
  {Abernathy}, {Acernese}, {Allen}, {Andersson}, {Arun}, {Barone}, {Barr},
  {Barsuglia}, {Beker} et~al.}}]{2010CQGra..27h4007P}
\bibinfo{author}{\bibfnamefont{M.}~\bibnamefont{{Punturo}}},
  \bibinfo{author}{\bibfnamefont{M.}~\bibnamefont{{Abernathy}}},
  \bibinfo{author}{\bibfnamefont{F.}~\bibnamefont{{Acernese}}},
  \bibinfo{author}{\bibfnamefont{B.}~\bibnamefont{{Allen}}},
  \bibinfo{author}{\bibfnamefont{N.}~\bibnamefont{{Andersson}}},
  \bibinfo{author}{\bibfnamefont{K.}~\bibnamefont{{Arun}}},
  \bibinfo{author}{\bibfnamefont{F.}~\bibnamefont{{Barone}}},
  \bibinfo{author}{\bibfnamefont{B.}~\bibnamefont{{Barr}}},
  \bibinfo{author}{\bibfnamefont{M.}~\bibnamefont{{Barsuglia}}},
  \bibinfo{author}{\bibfnamefont{M.}~\bibnamefont{{Beker}}},
  \bibnamefont{et~al.}, \bibinfo{journal}{Classical and Quantum Gravity}
  \textbf{\bibinfo{volume}{27}}, \bibinfo{eid}{084007} (\bibinfo{year}{2010}).

\bibitem[{\citenamefont{{Messenger} et~al.}(2013)\citenamefont{{Messenger},
  {Takami}, {Gossan}, {Rezzolla}, and {Sathyaprakash}}}]{2013arXiv1312.1862M}
\bibinfo{author}{\bibfnamefont{C.}~\bibnamefont{{Messenger}}},
  \bibinfo{author}{\bibfnamefont{K.}~\bibnamefont{{Takami}}},
  \bibinfo{author}{\bibfnamefont{S.}~\bibnamefont{{Gossan}}},
  \bibinfo{author}{\bibfnamefont{L.}~\bibnamefont{{Rezzolla}}},
  \bibnamefont{and} \bibinfo{author}{\bibfnamefont{B.~S.}
  \bibnamefont{{Sathyaprakash}}}, \bibinfo{journal}{ArXiv e-prints}
  (\bibinfo{year}{2013}), \eprint{1312.1862}.

\bibitem[{\citenamefont{{Schutz}}(1986)}]{1986Natur.323..310S}
\bibinfo{author}{\bibfnamefont{B.~F.} \bibnamefont{{Schutz}}},
  \bibinfo{journal}{\nat} \textbf{\bibinfo{volume}{323}}, \bibinfo{pages}{310}
  (\bibinfo{year}{1986}).

\bibitem[{\citenamefont{{Read} et~al.}(2013)\citenamefont{{Read}, {Baiotti},
  {Creighton}, {Friedman}, {Giacomazzo}, {Kyutoku}, {Markakis}, {Rezzolla},
  {Shibata}, and {Taniguchi}}}]{2013PhRvD..88d4042R}
\bibinfo{author}{\bibfnamefont{J.~S.} \bibnamefont{{Read}}},
  \bibinfo{author}{\bibfnamefont{L.}~\bibnamefont{{Baiotti}}},
  \bibinfo{author}{\bibfnamefont{J.~D.~E.} \bibnamefont{{Creighton}}},
  \bibinfo{author}{\bibfnamefont{J.~L.} \bibnamefont{{Friedman}}},
  \bibinfo{author}{\bibfnamefont{B.}~\bibnamefont{{Giacomazzo}}},
  \bibinfo{author}{\bibfnamefont{K.}~\bibnamefont{{Kyutoku}}},
  \bibinfo{author}{\bibfnamefont{C.}~\bibnamefont{{Markakis}}},
  \bibinfo{author}{\bibfnamefont{L.}~\bibnamefont{{Rezzolla}}},
  \bibinfo{author}{\bibfnamefont{M.}~\bibnamefont{{Shibata}}},
  \bibnamefont{and}
  \bibinfo{author}{\bibfnamefont{K.}~\bibnamefont{{Taniguchi}}},
  \bibinfo{journal}{\prd} \textbf{\bibinfo{volume}{88}}, \bibinfo{eid}{044042}
  (\bibinfo{year}{2013}), \eprint{1306.4065}.

\bibitem[{\citenamefont{{Anderson} et~al.}(2001)\citenamefont{{Anderson},
  {Brady}, {Creighton}, and {Flanagan}}}]{2001PhRvD..63d2003A}
\bibinfo{author}{\bibfnamefont{W.~G.} \bibnamefont{{Anderson}}},
  \bibinfo{author}{\bibfnamefont{P.~R.} \bibnamefont{{Brady}}},
  \bibinfo{author}{\bibfnamefont{J.~D.} \bibnamefont{{Creighton}}},
  \bibnamefont{and} \bibinfo{author}{\bibfnamefont{{\'E}.~{\'E}.}
  \bibnamefont{{Flanagan}}}, \bibinfo{journal}{\prd}
  \textbf{\bibinfo{volume}{63}}, \bibinfo{eid}{042003} (\bibinfo{year}{2001}),
  \eprint{gr-qc/0008066}.

\bibitem[{\citenamefont{{Chatterji} et~al.}(2004)\citenamefont{{Chatterji},
  {Blackburn}, {Martin}, and {Katsavounidis}}}]{2004CQGra..21S1809C}
\bibinfo{author}{\bibfnamefont{S.}~\bibnamefont{{Chatterji}}},
  \bibinfo{author}{\bibfnamefont{L.}~\bibnamefont{{Blackburn}}},
  \bibinfo{author}{\bibfnamefont{G.}~\bibnamefont{{Martin}}}, \bibnamefont{and}
  \bibinfo{author}{\bibfnamefont{E.}~\bibnamefont{{Katsavounidis}}},
  \bibinfo{journal}{Classical and Quantum Gravity}
  \textbf{\bibinfo{volume}{21}}, \bibinfo{pages}{1809} (\bibinfo{year}{2004}),
  \eprint{gr-qc/0412119}.

\bibitem[{\citenamefont{{Klimenko} and
  {Mitselmakher}}(2004)}]{2004CQGra..21S1819K}
\bibinfo{author}{\bibfnamefont{S.}~\bibnamefont{{Klimenko}}} \bibnamefont{and}
  \bibinfo{author}{\bibfnamefont{G.}~\bibnamefont{{Mitselmakher}}},
  \bibinfo{journal}{Classical and Quantum Gravity}
  \textbf{\bibinfo{volume}{21}}, \bibinfo{pages}{1819} (\bibinfo{year}{2004}).

\bibitem[{\citenamefont{{Sutton} et~al.}(2010)\citenamefont{{Sutton}, {Jones},
  {Chatterji}, {Kalmus}, {Leonor}, {Poprocki}, {Rollins}, {Searle}, {Stein},
  {Tinto} et~al.}}]{2010NJPh...12e3034S}
\bibinfo{author}{\bibfnamefont{P.~J.} \bibnamefont{{Sutton}}},
  \bibinfo{author}{\bibfnamefont{G.}~\bibnamefont{{Jones}}},
  \bibinfo{author}{\bibfnamefont{S.}~\bibnamefont{{Chatterji}}},
  \bibinfo{author}{\bibfnamefont{P.}~\bibnamefont{{Kalmus}}},
  \bibinfo{author}{\bibfnamefont{I.}~\bibnamefont{{Leonor}}},
  \bibinfo{author}{\bibfnamefont{S.}~\bibnamefont{{Poprocki}}},
  \bibinfo{author}{\bibfnamefont{J.}~\bibnamefont{{Rollins}}},
  \bibinfo{author}{\bibfnamefont{A.}~\bibnamefont{{Searle}}},
  \bibinfo{author}{\bibfnamefont{L.}~\bibnamefont{{Stein}}},
  \bibinfo{author}{\bibfnamefont{M.}~\bibnamefont{{Tinto}}},
  \bibnamefont{et~al.}, \bibinfo{journal}{New Journal of Physics}
  \textbf{\bibinfo{volume}{12}}, \bibinfo{eid}{053034} (\bibinfo{year}{2010}),
  \eprint{0908.3665}.

\bibitem[{\citenamefont{Klimenko et~al.}(2005)\citenamefont{Klimenko, Mohanty,
  Rakhmanov, and Mitselmakher}}]{Klimenko:2005wa}
\bibinfo{author}{\bibfnamefont{S.}~\bibnamefont{Klimenko}},
  \bibinfo{author}{\bibfnamefont{S.}~\bibnamefont{Mohanty}},
  \bibinfo{author}{\bibfnamefont{M.}~\bibnamefont{Rakhmanov}},
  \bibnamefont{and}
  \bibinfo{author}{\bibfnamefont{G.}~\bibnamefont{Mitselmakher}},
  \bibinfo{journal}{Physical Review D} \textbf{\bibinfo{volume}{72}},
  \bibinfo{pages}{122002} (\bibinfo{year}{2005}).

\bibitem[{\citenamefont{Klimenko et~al.}(2008)\citenamefont{Klimenko, Yakushin,
  Mercer, and Mitselmakher}}]{Klimenko:2007hd}
\bibinfo{author}{\bibfnamefont{S.}~\bibnamefont{Klimenko}},
  \bibinfo{author}{\bibfnamefont{I.}~\bibnamefont{Yakushin}},
  \bibinfo{author}{\bibfnamefont{A.}~\bibnamefont{Mercer}}, \bibnamefont{and}
  \bibinfo{author}{\bibfnamefont{G.}~\bibnamefont{Mitselmakher}},
  \bibinfo{journal}{Class. Quant. Grav.} \textbf{\bibinfo{volume}{25}},
  \bibinfo{pages}{114029} (\bibinfo{year}{2008}).

\bibitem[{\citenamefont{{Summerscales}
  et~al.}(2008)\citenamefont{{Summerscales}, {Burrows}, {Finn}, and
  {Ott}}}]{2008ApJ...678.1142S}
\bibinfo{author}{\bibfnamefont{T.~Z.} \bibnamefont{{Summerscales}}},
  \bibinfo{author}{\bibfnamefont{A.}~\bibnamefont{{Burrows}}},
  \bibinfo{author}{\bibfnamefont{L.~S.} \bibnamefont{{Finn}}},
  \bibnamefont{and} \bibinfo{author}{\bibfnamefont{C.~D.} \bibnamefont{{Ott}}},
  \bibinfo{journal}{\apj} \textbf{\bibinfo{volume}{678}}, \bibinfo{pages}{1142}
  (\bibinfo{year}{2008}), \eprint{0704.2157}.

\bibitem[{\citenamefont{{R{\"o}ver} et~al.}(2009)\citenamefont{{R{\"o}ver},
  {Bizouard}, {Christensen}, {Dimmelmeier}, {Heng}, and
  {Meyer}}}]{2009PhRvD..80j2004R}
\bibinfo{author}{\bibfnamefont{C.}~\bibnamefont{{R{\"o}ver}}},
  \bibinfo{author}{\bibfnamefont{M.-A.} \bibnamefont{{Bizouard}}},
  \bibinfo{author}{\bibfnamefont{N.}~\bibnamefont{{Christensen}}},
  \bibinfo{author}{\bibfnamefont{H.}~\bibnamefont{{Dimmelmeier}}},
  \bibinfo{author}{\bibfnamefont{I.~S.} \bibnamefont{{Heng}}},
  \bibnamefont{and} \bibinfo{author}{\bibfnamefont{R.}~\bibnamefont{{Meyer}}},
  \bibinfo{journal}{\prd} \textbf{\bibinfo{volume}{80}}, \bibinfo{eid}{102004}
  (\bibinfo{year}{2009}), \eprint{0909.1093}.

\bibitem[{\citenamefont{{Logue} et~al.}(2012)\citenamefont{{Logue}, {Ott},
  {Heng}, {Kalmus}, and {Scargill}}}]{2012PhRvD..86d4023L}
\bibinfo{author}{\bibfnamefont{J.}~\bibnamefont{{Logue}}},
  \bibinfo{author}{\bibfnamefont{C.~D.} \bibnamefont{{Ott}}},
  \bibinfo{author}{\bibfnamefont{I.~S.} \bibnamefont{{Heng}}},
  \bibinfo{author}{\bibfnamefont{P.}~\bibnamefont{{Kalmus}}}, \bibnamefont{and}
  \bibinfo{author}{\bibfnamefont{J.~H.~C.} \bibnamefont{{Scargill}}},
  \bibinfo{journal}{\prd} \textbf{\bibinfo{volume}{86}}, \bibinfo{eid}{044023}
  (\bibinfo{year}{2012}), \eprint{1202.3256}.

\bibitem[{\citenamefont{{Arun} et~al.}(2005)\citenamefont{{Arun}, {Iyer},
  {Sathyaprakash}, and {Sundararajan}}}]{2005PhRvD..71h4008A}
\bibinfo{author}{\bibfnamefont{K.~G.} \bibnamefont{{Arun}}},
  \bibinfo{author}{\bibfnamefont{B.~R.} \bibnamefont{{Iyer}}},
  \bibinfo{author}{\bibfnamefont{B.~S.} \bibnamefont{{Sathyaprakash}}},
  \bibnamefont{and} \bibinfo{author}{\bibfnamefont{P.~A.}
  \bibnamefont{{Sundararajan}}}, \bibinfo{journal}{\prd}
  \textbf{\bibinfo{volume}{71}}, \bibinfo{eid}{084008} (\bibinfo{year}{2005}),
  \eprint{gr-qc/0411146}.

\bibitem[{\citenamefont{{Fairhurst}}(2011)}]{2011CQGra..28j5021F}
\bibinfo{author}{\bibfnamefont{S.}~\bibnamefont{{Fairhurst}}},
  \bibinfo{journal}{Classical and Quantum Gravity}
  \textbf{\bibinfo{volume}{28}}, \bibinfo{eid}{105021} (\bibinfo{year}{2011}),
  \eprint{1010.6192}.

\bibitem[{\citenamefont{{Abadie}
  et~al.}(2012{\natexlab{a}})}]{2012PhRvD..85l2007A}
\bibinfo{author}{\bibfnamefont{J.}~\bibnamefont{{Abadie}}} \bibnamefont{et~al.}
  (\bibinfo{collaboration}{{LIGO Scientific Collaboration} and {Virgo
  Collaboration}}), \bibinfo{journal}{\prd} \textbf{\bibinfo{volume}{85}},
  \bibinfo{eid}{122007} (\bibinfo{year}{2012}{\natexlab{a}}),
  \eprint{1202.2788}.

\bibitem[{\citenamefont{{The LIGO Scientific Collaboration} and {the Virgo
  Collaboration: J. Abadie, et al.}}(2012)}]{Abadie:2012aa}
\bibinfo{author}{\bibnamefont{{The LIGO Scientific Collaboration}}}
  \bibnamefont{and} \bibinfo{author}{\bibnamefont{{the Virgo Collaboration: J.
  Abadie, et al.}}}, \bibinfo{journal}{Physical Review D}
  \textbf{\bibinfo{volume}{85}} (\bibinfo{year}{2012}).

\bibitem[{\citenamefont{{Allen} et~al.}(2012)\citenamefont{{Allen}, {Anderson},
  {Brady}, {Brown}, and {Creighton}}}]{2012PhRvD..85l2006A}
\bibinfo{author}{\bibfnamefont{B.}~\bibnamefont{{Allen}}},
  \bibinfo{author}{\bibfnamefont{W.~G.} \bibnamefont{{Anderson}}},
  \bibinfo{author}{\bibfnamefont{P.~R.} \bibnamefont{{Brady}}},
  \bibinfo{author}{\bibfnamefont{D.~A.} \bibnamefont{{Brown}}},
  \bibnamefont{and} \bibinfo{author}{\bibfnamefont{J.~D.~E.}
  \bibnamefont{{Creighton}}}, \bibinfo{journal}{\prd}
  \textbf{\bibinfo{volume}{85}}, \bibinfo{eid}{122006} (\bibinfo{year}{2012}),
  \eprint{gr-qc/0509116}.

\bibitem[{\citenamefont{{Abadie}
  et~al.}(2012{\natexlab{b}})}]{2012PhRvD..85h2002A}
\bibinfo{author}{\bibfnamefont{J.}~\bibnamefont{{Abadie}}} \bibnamefont{et~al.}
  (\bibinfo{collaboration}{{LIGO Scientific Collaboration} and {Virgo
  Collaboration}}), \bibinfo{journal}{\prd} \textbf{\bibinfo{volume}{85}},
  \bibinfo{eid}{082002} (\bibinfo{year}{2012}{\natexlab{b}}),
  \eprint{1111.7314}.

\bibitem[{\citenamefont{{The LIGO Scientific Collaboration} and {the Virgo
  Collaboration: J. Aasi, et al.}}(2013)}]{Aasi:2013aa}
\bibinfo{author}{\bibnamefont{{The LIGO Scientific Collaboration}}}
  \bibnamefont{and} \bibinfo{author}{\bibnamefont{{the Virgo Collaboration: J.
  Aasi, et al.}}}, \bibinfo{journal}{Physical Review D}
  \textbf{\bibinfo{volume}{87}} (\bibinfo{year}{2013}).

\bibitem[{\citenamefont{Blanchet}(2014)}]{lrr-2014-2}
\bibinfo{author}{\bibfnamefont{L.}~\bibnamefont{Blanchet}},
  \bibinfo{journal}{Living Reviews in Relativity} \textbf{\bibinfo{volume}{17}}
  (\bibinfo{year}{2014}),
  \urlprefix\url{http://www.livingreviews.org/lrr-2014-2}.

\bibitem[{\citenamefont{{Buonanno} et~al.}(2007)\citenamefont{{Buonanno},
  {Pan}, {Baker}, {Centrella}, {Kelly}, {McWilliams}, and {van
  Meter}}}]{2007PhRvD..76j4049B}
\bibinfo{author}{\bibfnamefont{A.}~\bibnamefont{{Buonanno}}},
  \bibinfo{author}{\bibfnamefont{Y.}~\bibnamefont{{Pan}}},
  \bibinfo{author}{\bibfnamefont{J.~G.} \bibnamefont{{Baker}}},
  \bibinfo{author}{\bibfnamefont{J.}~\bibnamefont{{Centrella}}},
  \bibinfo{author}{\bibfnamefont{B.~J.} \bibnamefont{{Kelly}}},
  \bibinfo{author}{\bibfnamefont{S.~T.} \bibnamefont{{McWilliams}}},
  \bibnamefont{and} \bibinfo{author}{\bibfnamefont{J.~R.} \bibnamefont{{van
  Meter}}}, \bibinfo{journal}{\prd} \textbf{\bibinfo{volume}{76}},
  \bibinfo{eid}{104049} (\bibinfo{year}{2007}), \eprint{0706.3732}.

\bibitem[{\citenamefont{{Ajith} et~al.}(2011)\citenamefont{{Ajith}, {Hannam},
  {Husa}, {Chen}, {Br{\"u}gmann}, {Dorband}, {M{\"u}ller}, {Ohme}, {Pollney},
  {Reisswig} et~al.}}]{2011PhRvL.106x1101A}
\bibinfo{author}{\bibfnamefont{P.}~\bibnamefont{{Ajith}}},
  \bibinfo{author}{\bibfnamefont{M.}~\bibnamefont{{Hannam}}},
  \bibinfo{author}{\bibfnamefont{S.}~\bibnamefont{{Husa}}},
  \bibinfo{author}{\bibfnamefont{Y.}~\bibnamefont{{Chen}}},
  \bibinfo{author}{\bibfnamefont{B.}~\bibnamefont{{Br{\"u}gmann}}},
  \bibinfo{author}{\bibfnamefont{N.}~\bibnamefont{{Dorband}}},
  \bibinfo{author}{\bibfnamefont{D.}~\bibnamefont{{M{\"u}ller}}},
  \bibinfo{author}{\bibfnamefont{F.}~\bibnamefont{{Ohme}}},
  \bibinfo{author}{\bibfnamefont{D.}~\bibnamefont{{Pollney}}},
  \bibinfo{author}{\bibfnamefont{C.}~\bibnamefont{{Reisswig}}},
  \bibnamefont{et~al.}, \bibinfo{journal}{Physical Review Letters}
  \textbf{\bibinfo{volume}{106}}, \bibinfo{eid}{241101} (\bibinfo{year}{2011}),
  \eprint{0909.2867}.

\bibitem[{\citenamefont{{Echeverria}}(1989)}]{1989PhRvD..40.3194E}
\bibinfo{author}{\bibfnamefont{F.}~\bibnamefont{{Echeverria}}},
  \bibinfo{journal}{\prd} \textbf{\bibinfo{volume}{40}}, \bibinfo{pages}{3194}
  (\bibinfo{year}{1989}).

\bibitem[{\citenamefont{{Jaranowski} et~al.}(1998)\citenamefont{{Jaranowski},
  {Kr{\'o}lak}, and {Schutz}}}]{1998PhRvD..58f3001J}
\bibinfo{author}{\bibfnamefont{P.}~\bibnamefont{{Jaranowski}}},
  \bibinfo{author}{\bibfnamefont{A.}~\bibnamefont{{Kr{\'o}lak}}},
  \bibnamefont{and} \bibinfo{author}{\bibfnamefont{B.~F.}
  \bibnamefont{{Schutz}}}, \bibinfo{journal}{\prd}
  \textbf{\bibinfo{volume}{58}}, \bibinfo{eid}{063001} (\bibinfo{year}{1998}),
  \eprint{gr-qc/9804014}.

\bibitem[{\citenamefont{{Pankow} et~al.}(2009)\citenamefont{{Pankow},
  {Klimenko}, {Mitselmakher}, {Yakushin}, {Vedovato}, {Drago}, {Mercer}, and
  {Ajith}}}]{2009CQGra..26t4004P}
\bibinfo{author}{\bibfnamefont{C.}~\bibnamefont{{Pankow}}},
  \bibinfo{author}{\bibfnamefont{S.}~\bibnamefont{{Klimenko}}},
  \bibinfo{author}{\bibfnamefont{G.}~\bibnamefont{{Mitselmakher}}},
  \bibinfo{author}{\bibfnamefont{I.}~\bibnamefont{{Yakushin}}},
  \bibinfo{author}{\bibfnamefont{G.}~\bibnamefont{{Vedovato}}},
  \bibinfo{author}{\bibfnamefont{M.}~\bibnamefont{{Drago}}},
  \bibinfo{author}{\bibfnamefont{R.~A.} \bibnamefont{{Mercer}}},
  \bibnamefont{and} \bibinfo{author}{\bibfnamefont{P.}~\bibnamefont{{Ajith}}},
  \bibinfo{journal}{Classical and Quantum Gravity}
  \textbf{\bibinfo{volume}{26}}, \bibinfo{eid}{204004} (\bibinfo{year}{2009}),
  \eprint{0905.3120}.

\bibitem[{\citenamefont{Abbott et~al.}(2009{\natexlab{a}})}]{s5_allsky}
\bibinfo{author}{\bibfnamefont{B.~P.} \bibnamefont{Abbott}}
  \bibnamefont{et~al.} (\bibinfo{collaboration}{LIGO Scientific
  Collaboration}), \bibinfo{journal}{Phys. Rev. D}
  \textbf{\bibinfo{volume}{80}}, \bibinfo{pages}{102001}
  (\bibinfo{year}{2009}{\natexlab{a}}),
  \urlprefix\url{http://link.aps.org/doi/10.1103/PhysRevD.80.102001}.

\bibitem[{\citenamefont{{Finn} and {Chernoff}}(1993)}]{1993PhRvD..47.2198F}
\bibinfo{author}{\bibfnamefont{L.~S.} \bibnamefont{{Finn}}} \bibnamefont{and}
  \bibinfo{author}{\bibfnamefont{D.~F.} \bibnamefont{{Chernoff}}},
  \bibinfo{journal}{\prd} \textbf{\bibinfo{volume}{47}}, \bibinfo{pages}{2198}
  (\bibinfo{year}{1993}), \eprint{gr-qc/9301003}.

\bibitem[{\citenamefont{{Cutler} and {Flanagan}}(1994)}]{1994PhRvD..49.2658C}
\bibinfo{author}{\bibfnamefont{C.}~\bibnamefont{{Cutler}}} \bibnamefont{and}
  \bibinfo{author}{\bibfnamefont{{\'E}.~E.} \bibnamefont{{Flanagan}}},
  \bibinfo{journal}{\prd} \textbf{\bibinfo{volume}{49}}, \bibinfo{pages}{2658}
  (\bibinfo{year}{1994}), \eprint{gr-qc/9402014}.

\bibitem[{\citenamefont{{Jaranowski} et~al.}(1996)\citenamefont{{Jaranowski},
  {Kokkotas}, {Kr{\'o}lak}, and {Tsegas}}}]{1996CQGra..13.1279J}
\bibinfo{author}{\bibfnamefont{P.}~\bibnamefont{{Jaranowski}}},
  \bibinfo{author}{\bibfnamefont{K.~D.} \bibnamefont{{Kokkotas}}},
  \bibinfo{author}{\bibfnamefont{A.}~\bibnamefont{{Kr{\'o}lak}}},
  \bibnamefont{and} \bibinfo{author}{\bibfnamefont{G.}~\bibnamefont{{Tsegas}}},
  \bibinfo{journal}{Classical and Quantum Gravity}
  \textbf{\bibinfo{volume}{13}}, \bibinfo{pages}{1279} (\bibinfo{year}{1996}).

\bibitem[{\citenamefont{{Veitch} et~al.}(2012)\citenamefont{{Veitch}, {Mandel},
  {Aylott}, {Farr}, {Raymond}, {Rodriguez}, {van der Sluys}, {Kalogera}, and
  {Vecchio}}}]{2012PhRvD..85j4045V}
\bibinfo{author}{\bibfnamefont{J.}~\bibnamefont{{Veitch}}},
  \bibinfo{author}{\bibfnamefont{I.}~\bibnamefont{{Mandel}}},
  \bibinfo{author}{\bibfnamefont{B.}~\bibnamefont{{Aylott}}},
  \bibinfo{author}{\bibfnamefont{B.}~\bibnamefont{{Farr}}},
  \bibinfo{author}{\bibfnamefont{V.}~\bibnamefont{{Raymond}}},
  \bibinfo{author}{\bibfnamefont{C.}~\bibnamefont{{Rodriguez}}},
  \bibinfo{author}{\bibfnamefont{M.}~\bibnamefont{{van der Sluys}}},
  \bibinfo{author}{\bibfnamefont{V.}~\bibnamefont{{Kalogera}}},
  \bibnamefont{and}
  \bibinfo{author}{\bibfnamefont{A.}~\bibnamefont{{Vecchio}}},
  \bibinfo{journal}{\prd} \textbf{\bibinfo{volume}{85}}, \bibinfo{eid}{104045}
  (\bibinfo{year}{2012}), \eprint{1201.1195}.

\bibitem[{\citenamefont{{Hannam} et~al.}(2013)\citenamefont{{Hannam}, {Brown},
  {Fairhurst}, {Fryer}, and {Harry}}}]{2013ApJ...766L..14H}
\bibinfo{author}{\bibfnamefont{M.}~\bibnamefont{{Hannam}}},
  \bibinfo{author}{\bibfnamefont{D.~A.} \bibnamefont{{Brown}}},
  \bibinfo{author}{\bibfnamefont{S.}~\bibnamefont{{Fairhurst}}},
  \bibinfo{author}{\bibfnamefont{C.~L.} \bibnamefont{{Fryer}}},
  \bibnamefont{and} \bibinfo{author}{\bibfnamefont{I.~W.}
  \bibnamefont{{Harry}}}, \bibinfo{journal}{Astrophys. J. Lett.}
  \textbf{\bibinfo{volume}{766}}, \bibinfo{eid}{L14} (\bibinfo{year}{2013}),
  \eprint{1301.5616}.

\bibitem[{\citenamefont{{Aasi}
  et~al.}(2013{\natexlab{b}})}]{2013PhRvD..88f2001A}
\bibinfo{author}{\bibfnamefont{J.}~\bibnamefont{{Aasi}}} \bibnamefont{et~al.}
  (\bibinfo{collaboration}{{LIGO Scientific Collaboration} and {Virgo
  Collaboration}}), \bibinfo{journal}{\prd} \textbf{\bibinfo{volume}{88}},
  \bibinfo{eid}{062001} (\bibinfo{year}{2013}{\natexlab{b}}),
  \eprint{1304.1775}.

\bibitem[{\citenamefont{{Schwarz}}(1978)}]{BIC_Schwarz}
\bibinfo{author}{\bibfnamefont{G.}~\bibnamefont{{Schwarz}}},
  \bibinfo{journal}{Annals of Statistics} \textbf{\bibinfo{volume}{6}},
  \bibinfo{pages}{461} (\bibinfo{year}{1978}).

\bibitem[{\citenamefont{Aasi et~al.}(2014)}]{ninja2}
\bibinfo{author}{\bibfnamefont{J.}~\bibnamefont{Aasi}} \bibnamefont{et~al.}
  (\bibinfo{collaboration}{The LIGO Scientific Collaboration, The Virgo
  Collaboration and The NINJA-2 Collaboration}), \bibinfo{journal}{Class.
  Quant. Grav.} \textbf{\bibinfo{volume}{31}}, \bibinfo{pages}{115004}
  (\bibinfo{year}{2014}), \eprint{1401.0939}.

\bibitem[{\citenamefont{{The LIGO Scientific
  Collaboration}}(2009)}]{aLIGO_noise_curves}
\bibinfo{author}{\bibnamefont{{The LIGO Scientific Collaboration}}}
  (\bibinfo{year}{2009}),
  \urlprefix\url{https://dcc.ligo.org/LIGO-T0900288/public}.

\bibitem[{\citenamefont{{Accadia} et~al.}(2012)}]{adV_noise_curves}
\bibinfo{author}{\bibfnamefont{T.}~\bibnamefont{{Accadia}}}
  \bibnamefont{et~al.} (\bibinfo{year}{2012}),
  \urlprefix\url{https://tds.ego-gw.it/ql/?c=8940}.

\bibitem[{\citenamefont{Abbott et~al.}(2009{\natexlab{b}})}]{s5_highfreq}
\bibinfo{author}{\bibfnamefont{B.~P.} \bibnamefont{Abbott}}
  \bibnamefont{et~al.} (\bibinfo{collaboration}{The LIGO Scientific
  Collaboration}), \bibinfo{journal}{Phys. Rev. D}
  \textbf{\bibinfo{volume}{80}}, \bibinfo{pages}{102002}
  (\bibinfo{year}{2009}{\natexlab{b}}),
  \urlprefix\url{http://link.aps.org/doi/10.1103/PhysRevD.80.102002}.

\bibitem[{\citenamefont{Abadie et~al.}(2010)}]{s52vsr1_allsky}
\bibinfo{author}{\bibfnamefont{J.}~\bibnamefont{Abadie}} \bibnamefont{et~al.}
  (\bibinfo{collaboration}{The LIGO Scientific Collaboration and The Virgo
  Collaboration}), \bibinfo{journal}{Phys. Rev. D}
  \textbf{\bibinfo{volume}{81}}, \bibinfo{pages}{102001}
  (\bibinfo{year}{2010}),
  \urlprefix\url{http://link.aps.org/doi/10.1103/PhysRevD.81.102001}.

\bibitem[{\citenamefont{{Isenberg} and {Nester}}(1980)}]{1980grg..conf...23I}
\bibinfo{author}{\bibfnamefont{J.}~\bibnamefont{{Isenberg}}} \bibnamefont{and}
  \bibinfo{author}{\bibfnamefont{J.}~\bibnamefont{{Nester}}}, in
  \emph{\bibinfo{booktitle}{General Relativity and Gravitation}}
  (\bibinfo{publisher}{Plenum Press, New York}, \bibinfo{year}{1980}),
  p.~\bibinfo{pages}{23}.

\bibitem[{\citenamefont{{Wilson} et~al.}(1996)\citenamefont{{Wilson},
  {Mathews}, and {Marronetti}}}]{1996PhRvD..54.1317W}
\bibinfo{author}{\bibfnamefont{J.~R.} \bibnamefont{{Wilson}}},
  \bibinfo{author}{\bibfnamefont{G.~J.} \bibnamefont{{Mathews}}},
  \bibnamefont{and}
  \bibinfo{author}{\bibfnamefont{P.}~\bibnamefont{{Marronetti}}},
  \bibinfo{journal}{\prd} \textbf{\bibinfo{volume}{54}}, \bibinfo{pages}{1317}
  (\bibinfo{year}{1996}).

\bibitem[{\citenamefont{{Oechslin} et~al.}(2002)\citenamefont{{Oechslin},
  {Rosswog}, and {Thielemann}}}]{2002PhRvD..65j3005O}
\bibinfo{author}{\bibfnamefont{R.}~\bibnamefont{{Oechslin}}},
  \bibinfo{author}{\bibfnamefont{S.}~\bibnamefont{{Rosswog}}},
  \bibnamefont{and} \bibinfo{author}{\bibfnamefont{F.-K.}
  \bibnamefont{{Thielemann}}}, \bibinfo{journal}{\prd}
  \textbf{\bibinfo{volume}{65}}, \bibinfo{pages}{103005}
  (\bibinfo{year}{2002}).

\bibitem[{\citenamefont{{Bauswein} et~al.}(2010)\citenamefont{{Bauswein},
  {Janka}, and {Oechslin}}}]{2010PhRvD..82h4043B}
\bibinfo{author}{\bibfnamefont{A.}~\bibnamefont{{Bauswein}}},
  \bibinfo{author}{\bibfnamefont{H.-T.} \bibnamefont{{Janka}}},
  \bibnamefont{and}
  \bibinfo{author}{\bibfnamefont{R.}~\bibnamefont{{Oechslin}}},
  \bibinfo{journal}{\prd} \textbf{\bibinfo{volume}{82}},
  \bibinfo{pages}{084043} (\bibinfo{year}{2010}).

\bibitem[{\citenamefont{{Balsara}}(1995)}]{1995JCoPh.121..357B}
\bibinfo{author}{\bibfnamefont{D.~S.} \bibnamefont{{Balsara}}},
  \bibinfo{journal}{Journal of Computational Physics}
  \textbf{\bibinfo{volume}{121}}, \bibinfo{pages}{357} (\bibinfo{year}{1995}).

\bibitem[{\citenamefont{{Lattimer}}(2012)}]{2012ARNPS..62..485L}
\bibinfo{author}{\bibfnamefont{J.~M.} \bibnamefont{{Lattimer}}},
  \bibinfo{journal}{Annu. Rev. Nucl. Part. Sci.} \textbf{\bibinfo{volume}{62}},
  \bibinfo{pages}{485} (\bibinfo{year}{2012}).

\bibitem[{\citenamefont{{Demorest} et~al.}(2010)\citenamefont{{Demorest},
  {Pennucci}, {Ransom}, {Roberts}, and {Hessels}}}]{2010Natur.467.1081D}
\bibinfo{author}{\bibfnamefont{P.~B.} \bibnamefont{{Demorest}}},
  \bibinfo{author}{\bibfnamefont{T.}~\bibnamefont{{Pennucci}}},
  \bibinfo{author}{\bibfnamefont{S.~M.} \bibnamefont{{Ransom}}},
  \bibinfo{author}{\bibfnamefont{M.~S.~E.} \bibnamefont{{Roberts}}},
  \bibnamefont{and} \bibinfo{author}{\bibfnamefont{J.~W.~T.}
  \bibnamefont{{Hessels}}}, \bibinfo{journal}{\nat}
  \textbf{\bibinfo{volume}{467}}, \bibinfo{pages}{1081} (\bibinfo{year}{2010}).

\bibitem[{\citenamefont{Antoniadis et~al.}(2013)\citenamefont{Antoniadis,
  Freire, Wex, Tauris, Lynch, van Kerkwijk, Kramer, Bassa, Dhillon, Driebe
  et~al.}}]{Antoniadis26042013}
\bibinfo{author}{\bibfnamefont{J.}~\bibnamefont{Antoniadis}},
  \bibinfo{author}{\bibfnamefont{P.~C.~C.} \bibnamefont{Freire}},
  \bibinfo{author}{\bibfnamefont{N.}~\bibnamefont{Wex}},
  \bibinfo{author}{\bibfnamefont{T.~M.} \bibnamefont{Tauris}},
  \bibinfo{author}{\bibfnamefont{R.~S.} \bibnamefont{Lynch}},
  \bibinfo{author}{\bibfnamefont{M.~H.} \bibnamefont{van Kerkwijk}},
  \bibinfo{author}{\bibfnamefont{M.}~\bibnamefont{Kramer}},
  \bibinfo{author}{\bibfnamefont{C.}~\bibnamefont{Bassa}},
  \bibinfo{author}{\bibfnamefont{V.~S.} \bibnamefont{Dhillon}},
  \bibinfo{author}{\bibfnamefont{T.}~\bibnamefont{Driebe}},
  \bibnamefont{et~al.}, \bibinfo{journal}{Science}
  \textbf{\bibinfo{volume}{340}} (\bibinfo{year}{2013}).

\bibitem[{\citenamefont{{Akmal} et~al.}(1998)\citenamefont{{Akmal},
  {Pandharipande}, and {Ravenhall}}}]{1998PhRvC..58.1804A}
\bibinfo{author}{\bibfnamefont{A.}~\bibnamefont{{Akmal}}},
  \bibinfo{author}{\bibfnamefont{V.~R.} \bibnamefont{{Pandharipande}}},
  \bibnamefont{and} \bibinfo{author}{\bibfnamefont{D.~G.}
  \bibnamefont{{Ravenhall}}}, \bibinfo{journal}{\prc}
  \textbf{\bibinfo{volume}{58}}, \bibinfo{pages}{1804} (\bibinfo{year}{1998}),
  \eprint{nucl-th/9804027}.

\bibitem[{\citenamefont{{Lalazissis} et~al.}(1997)\citenamefont{{Lalazissis},
  {K{\"o}nig}, and {Ring}}}]{1997PhRvC..55..540L}
\bibinfo{author}{\bibfnamefont{G.~A.} \bibnamefont{{Lalazissis}}},
  \bibinfo{author}{\bibfnamefont{J.}~\bibnamefont{{K{\"o}nig}}},
  \bibnamefont{and} \bibinfo{author}{\bibfnamefont{P.}~\bibnamefont{{Ring}}},
  \bibinfo{journal}{\prc} \textbf{\bibinfo{volume}{55}}, \bibinfo{pages}{540}
  (\bibinfo{year}{1997}).

\bibitem[{\citenamefont{{Hempel} and
  {Schaffner-Bielich}}(2010)}]{2010NuPhA.837..210H}
\bibinfo{author}{\bibfnamefont{M.}~\bibnamefont{{Hempel}}} \bibnamefont{and}
  \bibinfo{author}{\bibfnamefont{J.}~\bibnamefont{{Schaffner-Bielich}}},
  \bibinfo{journal}{Nucl. Phys. A} \textbf{\bibinfo{volume}{837}},
  \bibinfo{pages}{210} (\bibinfo{year}{2010}).

\bibitem[{\citenamefont{{Typel} et~al.}(2010)\citenamefont{{Typel},
  {R{\"o}pke}, {Kl{\"a}hn}, {Blaschke}, and {Wolter}}}]{2010PhRvC..81a5803T}
\bibinfo{author}{\bibfnamefont{S.}~\bibnamefont{{Typel}}},
  \bibinfo{author}{\bibfnamefont{G.}~\bibnamefont{{R{\"o}pke}}},
  \bibinfo{author}{\bibfnamefont{T.}~\bibnamefont{{Kl{\"a}hn}}},
  \bibinfo{author}{\bibfnamefont{D.}~\bibnamefont{{Blaschke}}},
  \bibnamefont{and} \bibinfo{author}{\bibfnamefont{H.~H.}
  \bibnamefont{{Wolter}}}, \bibinfo{journal}{\prc}
  \textbf{\bibinfo{volume}{81}}, \bibinfo{eid}{015803} (\bibinfo{year}{2010}).

\bibitem[{\citenamefont{{Shen} et~al.}(1998)\citenamefont{{Shen}, {Toki},
  {Oyamatsu}, and {Sumiyoshi}}}]{1998NuPhA.637..435S}
\bibinfo{author}{\bibfnamefont{H.}~\bibnamefont{{Shen}}},
  \bibinfo{author}{\bibfnamefont{H.}~\bibnamefont{{Toki}}},
  \bibinfo{author}{\bibfnamefont{K.}~\bibnamefont{{Oyamatsu}}},
  \bibnamefont{and}
  \bibinfo{author}{\bibfnamefont{K.}~\bibnamefont{{Sumiyoshi}}},
  \bibinfo{journal}{Nucl. Phys. A} \textbf{\bibinfo{volume}{637}},
  \bibinfo{pages}{435} (\bibinfo{year}{1998}).

\bibitem[{\citenamefont{{Sugahara} and {Toki}}(1994)}]{1994NuPhA.579..557S}
\bibinfo{author}{\bibfnamefont{Y.}~\bibnamefont{{Sugahara}}} \bibnamefont{and}
  \bibinfo{author}{\bibfnamefont{H.}~\bibnamefont{{Toki}}},
  \bibinfo{journal}{Nuclear Physics A} \textbf{\bibinfo{volume}{579}},
  \bibinfo{pages}{557} (\bibinfo{year}{1994}).

\bibitem[{\citenamefont{{Hempel} et~al.}(2012)\citenamefont{{Hempel},
  {Fischer}, {Schaffner-Bielich}, and
  {Liebend{\"o}rfer}}}]{2012ApJ...748...70H}
\bibinfo{author}{\bibfnamefont{M.}~\bibnamefont{{Hempel}}},
  \bibinfo{author}{\bibfnamefont{T.}~\bibnamefont{{Fischer}}},
  \bibinfo{author}{\bibfnamefont{J.}~\bibnamefont{{Schaffner-Bielich}}},
  \bibnamefont{and}
  \bibinfo{author}{\bibfnamefont{M.}~\bibnamefont{{Liebend{\"o}rfer}}},
  \bibinfo{journal}{\apj} \textbf{\bibinfo{volume}{748}}, \bibinfo{eid}{70}
  (\bibinfo{year}{2012}).

\bibitem[{\citenamefont{{Steiner} et~al.}(2013)\citenamefont{{Steiner},
  {Hempel}, and {Fischer}}}]{2013ApJ...774...17S}
\bibinfo{author}{\bibfnamefont{A.~W.} \bibnamefont{{Steiner}}},
  \bibinfo{author}{\bibfnamefont{M.}~\bibnamefont{{Hempel}}}, \bibnamefont{and}
  \bibinfo{author}{\bibfnamefont{T.}~\bibnamefont{{Fischer}}},
  \bibinfo{journal}{Astrophys. J.} \textbf{\bibinfo{volume}{774}},
  \bibinfo{eid}{17} (\bibinfo{year}{2013}), \eprint{1207.2184}.

\bibitem[{\citenamefont{{Toki} et~al.}(1995)\citenamefont{{Toki}, {Hirata},
  {Sugahara}, {Sumiyoshi}, and {Tanihata}}}]{1995NuPhA.588..357T}
\bibinfo{author}{\bibfnamefont{H.}~\bibnamefont{{Toki}}},
  \bibinfo{author}{\bibfnamefont{D.}~\bibnamefont{{Hirata}}},
  \bibinfo{author}{\bibfnamefont{Y.}~\bibnamefont{{Sugahara}}},
  \bibinfo{author}{\bibfnamefont{K.}~\bibnamefont{{Sumiyoshi}}},
  \bibnamefont{and}
  \bibinfo{author}{\bibfnamefont{I.}~\bibnamefont{{Tanihata}}},
  \bibinfo{journal}{Nuclear Physics A} \textbf{\bibinfo{volume}{588}},
  \bibinfo{pages}{357} (\bibinfo{year}{1995}).

\bibitem[{\citenamefont{{Bildsten} and {Cutler}}(1992)}]{1992ApJ...400..175B}
\bibinfo{author}{\bibfnamefont{L.}~\bibnamefont{{Bildsten}}} \bibnamefont{and}
  \bibinfo{author}{\bibfnamefont{C.}~\bibnamefont{{Cutler}}},
  \bibinfo{journal}{Astrophys. J.} \textbf{\bibinfo{volume}{400}},
  \bibinfo{pages}{175} (\bibinfo{year}{1992}).

\bibitem[{\citenamefont{{Kochanek}}(1992)}]{1992ApJ...398..234K}
\bibinfo{author}{\bibfnamefont{C.~S.} \bibnamefont{{Kochanek}}},
  \bibinfo{journal}{\apj} \textbf{\bibinfo{volume}{398}}, \bibinfo{pages}{234}
  (\bibinfo{year}{1992}).

\bibitem[{\citenamefont{{Dominik} et~al.}(2012)\citenamefont{{Dominik},
  {Belczynski}, {Fryer}, {Holz}, {Berti}, {Bulik}, {Mandel}, and
  {O'Shaughnessy}}}]{2012ApJ...759...52D}
\bibinfo{author}{\bibfnamefont{M.}~\bibnamefont{{Dominik}}},
  \bibinfo{author}{\bibfnamefont{K.}~\bibnamefont{{Belczynski}}},
  \bibinfo{author}{\bibfnamefont{C.}~\bibnamefont{{Fryer}}},
  \bibinfo{author}{\bibfnamefont{D.~E.} \bibnamefont{{Holz}}},
  \bibinfo{author}{\bibfnamefont{E.}~\bibnamefont{{Berti}}},
  \bibinfo{author}{\bibfnamefont{T.}~\bibnamefont{{Bulik}}},
  \bibinfo{author}{\bibfnamefont{I.}~\bibnamefont{{Mandel}}}, \bibnamefont{and}
  \bibinfo{author}{\bibfnamefont{R.}~\bibnamefont{{O'Shaughnessy}}},
  \bibinfo{journal}{\apj} \textbf{\bibinfo{volume}{759}}, \bibinfo{eid}{52}
  (\bibinfo{year}{2012}).

\bibitem[{\citenamefont{{Bauswein}
  et~al.}(2012{\natexlab{b}})\citenamefont{{Bauswein}, {Janka}, {Hebeler}, and
  {Schwenk}}}]{Bauswein38Eos}
\bibinfo{author}{\bibfnamefont{A.}~\bibnamefont{{Bauswein}}},
  \bibinfo{author}{\bibfnamefont{H.-T.} \bibnamefont{{Janka}}},
  \bibinfo{author}{\bibfnamefont{K.}~\bibnamefont{{Hebeler}}},
  \bibnamefont{and}
  \bibinfo{author}{\bibfnamefont{A.}~\bibnamefont{{Schwenk}}},
  \bibinfo{journal}{\prd} \textbf{\bibinfo{volume}{86}}, \bibinfo{eid}{063001}
  (\bibinfo{year}{2012}{\natexlab{b}}), \eprint{1204.1888}.

\bibitem[{\citenamefont{{Shibata} and {Taniguchi}}(2006)}]{2006PhRvD..73f4027S}
\bibinfo{author}{\bibfnamefont{M.}~\bibnamefont{{Shibata}}} \bibnamefont{and}
  \bibinfo{author}{\bibfnamefont{K.}~\bibnamefont{{Taniguchi}}},
  \bibinfo{journal}{\prd} \textbf{\bibinfo{volume}{73}}, \bibinfo{eid}{064027}
  (\bibinfo{year}{2006}), \eprint{astro-ph/0603145}.

\bibitem[{\citenamefont{{Shapiro}}(2000)}]{2000ApJ...544..397S}
\bibinfo{author}{\bibfnamefont{S.~L.} \bibnamefont{{Shapiro}}},
  \bibinfo{journal}{\apj} \textbf{\bibinfo{volume}{544}}, \bibinfo{pages}{397}
  (\bibinfo{year}{2000}), \eprint{astro-ph/0010493}.

\bibitem[{\citenamefont{{Paschalidis} et~al.}(2012)\citenamefont{{Paschalidis},
  {Etienne}, and {Shapiro}}}]{2012PhRvD..86f4032P}
\bibinfo{author}{\bibfnamefont{V.}~\bibnamefont{{Paschalidis}}},
  \bibinfo{author}{\bibfnamefont{Z.~B.} \bibnamefont{{Etienne}}},
  \bibnamefont{and} \bibinfo{author}{\bibfnamefont{S.~L.}
  \bibnamefont{{Shapiro}}}, \bibinfo{journal}{\prd}
  \textbf{\bibinfo{volume}{86}}, \bibinfo{eid}{064032} (\bibinfo{year}{2012}),
  \eprint{1208.5487}.

\bibitem[{\citenamefont{{Reisenegger} and
  {Bonacic}}(2003)}]{2003pasb.conf..231R}
\bibinfo{author}{\bibfnamefont{A.}~\bibnamefont{{Reisenegger}}}
  \bibnamefont{and}
  \bibinfo{author}{\bibfnamefont{A.}~\bibnamefont{{Bonacic}}}, in
  \emph{\bibinfo{booktitle}{Pulsars, AXPs and SGRs Observed with BeppoSAX and
  Other Observatories}}, edited by
  \bibinfo{editor}{\bibfnamefont{G.}~\bibnamefont{{Cusumano}}},
  \bibinfo{editor}{\bibfnamefont{E.}~\bibnamefont{{Massaro}}},
  \bibnamefont{and} \bibinfo{editor}{\bibfnamefont{T.}~\bibnamefont{{Mineo}}}
  (\bibinfo{year}{2003}), pp. \bibinfo{pages}{231--236},
  \eprint{astro-ph/0303454}.

\bibitem[{\citenamefont{{Andersson} and
  {Kokkotas}}(1998)}]{1998MNRAS.299.1059A}
\bibinfo{author}{\bibfnamefont{N.}~\bibnamefont{{Andersson}}} \bibnamefont{and}
  \bibinfo{author}{\bibfnamefont{K.~D.} \bibnamefont{{Kokkotas}}},
  \bibinfo{journal}{Mon. Not. Roy. Astron. Soc.}
  \textbf{\bibinfo{volume}{299}}, \bibinfo{pages}{1059} (\bibinfo{year}{1998}),
  \eprint{gr-qc/9711088}.

\bibitem[{\citenamefont{{John L. Friedman} and {Nikolaos
  Stergioulas}}(2013)}]{FriedmanStergioulas}
\bibinfo{author}{\bibnamefont{{John L. Friedman}}} \bibnamefont{and}
  \bibinfo{author}{\bibnamefont{{Nikolaos Stergioulas}}},
  \emph{\bibinfo{title}{Rotating Relativistic Stars}}
  (\bibinfo{publisher}{Cambridge University Press}, \bibinfo{year}{2013}).

\bibitem[{\citenamefont{{Gaertig} and {Kokkotas}}(2011)}]{2011PhRvD..83f4031G}
\bibinfo{author}{\bibfnamefont{E.}~\bibnamefont{{Gaertig}}} \bibnamefont{and}
  \bibinfo{author}{\bibfnamefont{K.~D.} \bibnamefont{{Kokkotas}}},
  \bibinfo{journal}{\prd} \textbf{\bibinfo{volume}{83}}, \bibinfo{eid}{064031}
  (\bibinfo{year}{2011}), \eprint{1005.5228}.

\bibitem[{\citenamefont{{Doneva} et~al.}(2013)\citenamefont{{Doneva},
  {Gaertig}, {Kokkotas}, and {Kr{\"u}ger}}}]{2013PhRvD..88d4052D}
\bibinfo{author}{\bibfnamefont{D.~D.} \bibnamefont{{Doneva}}},
  \bibinfo{author}{\bibfnamefont{E.}~\bibnamefont{{Gaertig}}},
  \bibinfo{author}{\bibfnamefont{K.~D.} \bibnamefont{{Kokkotas}}},
  \bibnamefont{and}
  \bibinfo{author}{\bibfnamefont{C.}~\bibnamefont{{Kr{\"u}ger}}},
  \bibinfo{journal}{\prd} \textbf{\bibinfo{volume}{88}}, \bibinfo{eid}{044052}
  (\bibinfo{year}{2013}), \eprint{1305.7197}.

\bibitem[{\citenamefont{{Zink} et~al.}(2010)\citenamefont{{Zink}, {Korobkin},
  {Schnetter}, and {Stergioulas}}}]{2010PhRvD..81h4055Z}
\bibinfo{author}{\bibfnamefont{B.}~\bibnamefont{{Zink}}},
  \bibinfo{author}{\bibfnamefont{O.}~\bibnamefont{{Korobkin}}},
  \bibinfo{author}{\bibfnamefont{E.}~\bibnamefont{{Schnetter}}},
  \bibnamefont{and}
  \bibinfo{author}{\bibfnamefont{N.}~\bibnamefont{{Stergioulas}}},
  \bibinfo{journal}{\prd} \textbf{\bibinfo{volume}{81}}, \bibinfo{eid}{084055}
  (\bibinfo{year}{2010}), \eprint{1003.0779}.

\bibitem[{\citenamefont{{Kr{\"u}ger} et~al.}(2010)\citenamefont{{Kr{\"u}ger},
  {Gaertig}, and {Kokkotas}}}]{2010PhRvD..81h4019K}
\bibinfo{author}{\bibfnamefont{C.}~\bibnamefont{{Kr{\"u}ger}}},
  \bibinfo{author}{\bibfnamefont{E.}~\bibnamefont{{Gaertig}}},
  \bibnamefont{and} \bibinfo{author}{\bibfnamefont{K.~D.}
  \bibnamefont{{Kokkotas}}}, \bibinfo{journal}{\prd}
  \textbf{\bibinfo{volume}{81}}, \bibinfo{eid}{084019} (\bibinfo{year}{2010}),
  \eprint{0911.2764}.

\bibitem[{\citenamefont{{Burgio} et~al.}(2011)\citenamefont{{Burgio},
  {Ferrari}, {Gualtieri}, and {Schulze}}}]{2011PhRvD..84d4017B}
\bibinfo{author}{\bibfnamefont{G.~F.} \bibnamefont{{Burgio}}},
  \bibinfo{author}{\bibfnamefont{V.}~\bibnamefont{{Ferrari}}},
  \bibinfo{author}{\bibfnamefont{L.}~\bibnamefont{{Gualtieri}}},
  \bibnamefont{and} \bibinfo{author}{\bibfnamefont{H.-J.}
  \bibnamefont{{Schulze}}}, \bibinfo{journal}{\prd}
  \textbf{\bibinfo{volume}{84}}, \bibinfo{eid}{044017} (\bibinfo{year}{2011}),
  \eprint{1106.2736}.

\bibitem[{\citenamefont{Jaranowski and Królak}(2012)}]{lrr-2012-4}
\bibinfo{author}{\bibfnamefont{P.}~\bibnamefont{Jaranowski}} \bibnamefont{and}
  \bibinfo{author}{\bibfnamefont{A.}~\bibnamefont{Królak}},
  \bibinfo{journal}{Living Reviews in Relativity} \textbf{\bibinfo{volume}{15}}
  (\bibinfo{year}{2012}),
  \urlprefix\url{http://www.livingreviews.org/lrr-2012-4}.

\bibitem[{\citenamefont{{Sutton}}(2013)}]{SuttonBursts}
\bibinfo{author}{\bibfnamefont{P.~J.} \bibnamefont{{Sutton}}},
  \bibinfo{journal}{ArXiv e-prints}  (\bibinfo{year}{2013}),
  \eprint{1304.0210}.

\bibitem[{\citenamefont{{Littenberg} and
  {Cornish}}(2010)}]{2010PhRvD..82j3007L}
\bibinfo{author}{\bibfnamefont{T.~B.} \bibnamefont{{Littenberg}}}
  \bibnamefont{and} \bibinfo{author}{\bibfnamefont{N.~J.}
  \bibnamefont{{Cornish}}}, \bibinfo{journal}{\prd}
  \textbf{\bibinfo{volume}{82}}, \bibinfo{eid}{103007} (\bibinfo{year}{2010}),
  \eprint{1008.1577}.

\bibitem[{\citenamefont{{Cornish}}(2012)}]{2012arXiv1204.2000C}
\bibinfo{author}{\bibfnamefont{N.~J.} \bibnamefont{{Cornish}}},
  \bibinfo{journal}{ArXiv e-prints}  (\bibinfo{year}{2012}),
  \eprint{1204.2000}.

\end{thebibliography}

\end{document}